\documentclass[twocolumn,superscriptaddress,showpacs,preprintnumbers,nofootinbib,amsmath,amssymb,floatfix,aps]{revtex4-1}
\usepackage{dcolumn}
\usepackage{bm}
\usepackage{color}

\usepackage{graphicx}
\usepackage{amsmath}
\usepackage{amssymb}
\usepackage{slashed}

\def\gsim{\buildrel > \over {_{\sim}}}

\def\beq{\begin{equation}}
\def\eeq{\end{equation}}
\def\be{\begin{eqnarray}}
\def\ee{\end{eqnarray}}
\def\ra{\rangle}
\def\la{\langle}

\begin{document}
\title{Scaling within the Spectral Function approach}
\author{J. E. Sobczyk}
\affiliation{Instituto de F\'\i sica Corpuscular (IFIC), Centro Mixto
CSIC-Universidad de Valencia, Institutos de Investigaci\'on de
Paterna, Apartado 22085, E-46071 Valencia, Spain}
\author{N. Rocco}
\affiliation{Department of Physics, University of Surrey, Guildford, GU2 7HX, UK}
\author{A. Lovato}
\affiliation{INFN-TIFPA Trento Institute of Fundamental Physics and Applications, 38123 Trento, Italy}
\affiliation{Physics Division, Argonne National Laboratory, Argonne, IL 60439}
\author{J. Nieves}
\affiliation{Instituto de F\'\i sica Corpuscular (IFIC), Centro Mixto
CSIC-Universidad de Valencia, Institutos de Investigaci\'on de
Paterna, Apartado 22085, E-46071 Valencia, Spain}

\date{\today}
\begin{abstract}
{Scaling features of the nuclear electromagnetic response functions unveil aspects of nuclear dynamics that are crucial
for interpretating neutrino- and electron-scattering data. In the large momentum-transfer regime, the nucleon-density 
response function defines a universal scaling function, which is independent of the nature of the probe. In this work, 
we analyze the nucleon-density response function of $^{12}$C, neglecting collective excitations. We employ particle and
hole spectral functions obtained within two distinct many-body methods, both widely used to describe electroweak 
reactions in nuclei. We show that the two approaches provide compatible nucleon-density scaling functions that for large
momentum transfers satisfy first-kind scaling. Both methods yield scaling functions characterized by an asymmetric shape,
although less pronounced than that of experimental scaling functions. This asymmetry, only mildly affected by final state 
interactions, is mostly due to nucleon-nucleon correlations, encoded in the continuum component of the hole SF.}

\end{abstract}
\maketitle

\section{Introduction}

\label{scaling}
The analysis of scaling properties of nuclear response functions has proven to be a useful tool to unveil 
information on the underlying nuclear structure and dynamics. Indeed, singling-out individual-nucleon interactions allows to 
disentangle the many-body aspects of the calculation. These properties are relevant for interpreting 
electron-scattering data and to predict quantities of interest for neutrino-oscillation experiments. It has been proposed 
that an empirical scaling function extracted from electron scattering data can be used to predict neutrino-nucleus
cross sections or to validate neutrino-nucleus interaction models~\cite{Caballero:2005sj, Antonov:2011bi, 
Amaro:2006tf, Gonzalez-Jimenez:2014eqa}. In particular, the use of relativistic mean field in such  calculations 
has found support in its capability of properly reproducing the asymmetric shape and 
the transverse enhancement of the empirical scaling function~\cite{Caballero:2005sj}. 

Recently, the authors of Ref.~\cite{Rocco:2017hmh} carried out an analysis of the scaling properties of the 
electromagnetic response functions of $^4$He and $^{12}$C nuclei computed by the Green's Function Monte 
Carlo (GFMC) approach~\cite{Carlson:2014vla}, retaining only one-body current contributions. Their results 
are consistent with scaling of  zeroth, first and second kind and show that the characteristic asymmetric shape 
of the experimental scaling function emerges in the calculations in spite of the non relativistic nature of the model. 
A novel interpretation of the longitudinal and transverse scaling functions in terms of a universal scaling function, 
defined in terms of the nucleon-density response function was discussed. However, the reason why the
nucleon-density scaling function depends on the energy and momentum transfers only through the scaling variable is yet 
to be fully understood. 

GFMC allows for a very accurate description of the properties of A $\leq 12$ nuclei, giving full
account of the dynamics of the constituent nucleons. However, within GFMC,  it is not straightforward to identify
the mechanisms responsible for the asymmetric shape of the scaling functions. In addition, only the leading 
relativistic corrections are included in the GFMC scaling functions, preventing a fully consistent comparison with 
the experimental ones. In fact, by employing both relativistic and non relativistic prefactors, it was possible to
highlight the shortcomings of GFMC in describing the electromagnetic responses at large momentum transfers~\cite{Rocco:2017hmh}.

In this work, we analyze the scaling properties of the electromagnetic responses in the moderate and
large momentum-transfer regions, where collective modes are unimportant and the spectral function (SF)
formalism is supposed to be reliable. This formalism, based on the impulse approximation (IA),
combines a fully relativistic description of the electromagnetic interaction with an accurate treatment of nuclear 
dynamics in the initial state. However, final state interactions (FSI) involving the struck particle are treated as 
corrections, whose inclusion requires further approximations~\cite{Benhar:2013dq,Ankowski:2014yfa}. 

Accurate calculations of the hole SF have been carried out in Refs.~\cite{Benhar:1989aw,Benhar:1994hw} within the correlated
basis function (CBF) theory. Being the struck nucleon relativistic, the particle SF
cannot be consistently derived within CBF, as the latter is an intrinsically non relativistic approach. Hence, FSI are usually
included by means of a convolution scheme.  The validity of this approximation has been recently tested by comparing SF and GFMC results for the one-body 
electromagnetic responses of $^{12}$C~\cite{Rocco:2016ejr}. The CBF-SF model has proven to successfully reproduce a large
body of electron scattering data for a variety of nuclear targets, up to relatively low momentum transfers, 
where the applicability of the IA is more controversial~\cite{Benhar:2006wy,Ankowski:2014yfa}. 
Recently, this model has been generalized to include the contributions of meson-exchange currents leading to 
final states with two  nucleons in the continuum~\cite{Benhar:2015ula,Rocco:2015cil}. The CBF-SF has
also been employed to describe neutrino-nucleus interactions~\cite{Benhar:2005dj, Benhar:2006nr, Benhar:2009wi, 
Benhar:2010nx, Vagnoni:2017hll} in both the quasiealstic and deep-inelastic scattering (DIS) regions.

In this work we also discuss a non relativistic semi-phenomenological approach, based on the local Fermi gas 
(LFG) model employed in Refs.~\cite{Nieves:2004wx,Nieves:2005rq} to study charge and neutral current quasielastic 
neutrino-nucleus scattering at intermediate and low energies. Within this model, 
the hole and the particle SFs are consistently derived in uniform and isospin-symmetric nuclear 
matter~\cite{FernandezdeCordoba:1991wf} and the local density approximation (LDA) is exploited to make
predictions for finite nuclear systems~\cite{Carrasco:1989vq, Nieves:1993ev, Gil:1997bm, GarciaRecio:2002cu, Nieves:2017lij,Sobczyk:2017mts}. We include relativistic corrections as 
in~\cite{Marco:1995vb,FernandezdeCordoba:1995pt} to extend the applicability of the model to moderately 
high momentum and energy transfers. We show that the particle spectral 
function can be employed to account for FSI with a comparable degree of accuracy as the convolution 
scheme.

In Sec.~\ref{sec:formalism} the scaling formalism is introduced; the LDA-based model allowing to consistently derive 
the hole and particle SFs is presented in Sec.~\ref{sec:lda};  Sec.~\ref{sec:ia} is devoted to
the CBF-SF approach and the inclusion of FSI. In Sec.~\ref{results}, the nucleon-density scaling
functions obtained within these two models are benchmarked and compared with those 
extracted from experimental data. In Sec.~\ref{sec:toy}, we discuss the origin of first-kind scaling,
and the asymmetry of the scaling function, employing  a simplified model for the nuclear dynamics. Finally,
in Sec.~\ref{sec:conclusions} we draw our conclusions.

\section{Scaling formalism}
\label{sec:formalism}
The electromagnetic longitudinal and transverse response functions are given by
\begin{align}
R_{\alpha}({\bf q},\omega)&=\sum_f \langle f | J_{\alpha}(\mathbf{q},\omega) | 0\rangle \langle 0| J^\dagger_{\alpha}(\mathbf{q},\omega) |f\rangle\nonumber\\
&\times \delta(\omega-E_f + E_0)\ ,
\label{eq:RLT}
\end{align}
where $|0\rangle$ and $|f\rangle$ represent the nuclear initial ground-state and final bound- or scattering-state of energies $E_0$
and $E_f$, respectively, and $J_{\alpha}(\mathbf{q},\omega)$ ($\alpha=L,T$) denotes the longitudinal and transverse components of the electromagnetic current.

The scaling properties of the nuclear responses have been widely analyzed in the framework of the Global Relativistic Fermi gas (GRFG) model. 
Within GRFG, the target nucleus is described as a collection of relativistic non-interacting nucleons, carrying a momentum smaller than the Fermi momentum $p_F$. In order to make contact with previous studies, we introduce the following set of dimensionless variables~\cite{Alberico:1988bv} 
\begin{align}
\lambda=&\omega/2m\ ,\nonumber\\
\kappa=&|{\bf q}|/2m\ ,\nonumber\\
\tau= &\kappa^2-\lambda^2\ ,\nonumber\\
\eta_F=&p_F/m\ ,\nonumber\\
\xi_F=& \frac{\sqrt{p_F^2+m^2}}{m}-1\ .\label{eq:defxi}
\end{align}
with $m$ the nucleon mass, and $q^\mu=(\omega, {\bf q})$ the four momentum transfer. 
A dimensionless scaling variable can be defined in terms of these quantities as~\cite{Alberico:1988bv}
\begin{align} 
\psi= \frac{1}{\sqrt{\xi_F}}\frac{\lambda-\tau}{\sqrt{(1+\lambda)\tau+\kappa\sqrt{\tau(1+\tau)}}} \,.
\end{align}

The longitudinal and transverse scaling functions are obtained by dividing the response functions by appropriate prefactors, 
encompassing single-nucleon dynamics within the GRFG model~\cite{Rocco:2017hmh}
\begin{equation}
f_{L,T}(\psi)=p_F\times \frac{R_{L,T}}{G_{L,T}}\, .
\label{eq:scal_func}
\end{equation}

It has to be noted that the GRFG longitudinal and transverse scaling functions coincide. The
analytical expression of the common function, symmetric and centered in $\psi=0$, reads
\begin{align}
f_L^{\rm GRFG}(\psi)= f_T^{\rm GRFG}(\psi)=\frac{3\xi_F}{2 \eta_F^2}\big(1-\psi^2)\theta(1-\psi^2)\ . \label{eq:fgrfg}
\end{align} 
The aim of our work is to discuss how the inclusion of nuclear interactions affects the shape of the scaling functions, possibly leading to scaling violations. 

In Ref.~\cite{Rocco:2017hmh} it has been suggested that, for large momentum transfers,  the longitudinal and transverse scaling functions can be interpreted in 
terms of the proton and neutron-density responses
\begin{align}
R_{p(n)}(\mathbf{q},\omega)=&\sum_f \langle  0| \varrho^\dagger_{p(n)}(\mathbf{q}) | f \rangle  \langle  f | \varrho_{p(n)}(\mathbf{q}) |   0 \rangle\, \nonumber\\
&\times\delta (\omega-E_f+E_0)   \ ,
\end{align}
where the proton (neutron)-density operator is given by
\begin{equation}
\varrho_{p(n)}(\mathbf{q})\equiv \sum_j e^{i \mathbf{q}\cdot \mathbf{r}_j} \frac{(1\pm\tau_{j,z})}{2}\, .
\end{equation} 
In isospin-symmetric nuclear matter, the proton- and neutron-density responses coincide. It is convenient to refer to them as nucleon-density response, proportional to the imaginary-part of the polarization propagator 
\begin{align}
S({\bf q},\omega)=\frac{1}{\pi}{\rm Im}\ \Pi({\bf q},\omega)\, ,
\label{dens:resp_def}
\end{align}
with
\begin{align}
\Pi({\bf q},\omega)= \la 0|\varrho^\dagger_{\bf q}\frac{1}{H-E_0-\omega-i\epsilon}\varrho_{\bf q}|0\rangle\ ,
\end{align}
where $H$ is the Hamiltonian, and  $\varrho_{\bf q}=\sum_{\bf p}a^\dagger_{\bf p+q}a_{\bf p}$
the proton- or neutron-density fluctuation operator. In the limit of large momentum transfer and for isospin symmetric nuclei,  
the nucleon-density scaling function $f$ is given by ~\cite{Rocco:2017hmh}
\begin{align}
f({\psi})= p_F\times 2\kappa\ S({\bf q},\omega) / \mathcal{N}
\label{scal:func}
\end{align} 
where ${\cal N}$ is either the number of protons or neutrons of the system. 

The one-body Green function in nuclear matter is defined as~\cite{fetterwalecka}, 
\begin{align}
G({\bf p},E)=& \langle 0| a^\dagger_{\bf p}\frac{1}{E+(H-E_0)-i\epsilon}a_{\bf p}|0\rangle\nonumber\\
& + \langle 0| a_{\bf p}\frac{1}{E-(H-E_0)+i\epsilon}a^\dagger_{\bf p}|0\rangle\nonumber\\
&= G_h({\bf p},E)+G_p({\bf p},E)\ .
\label{greens-fun}
\end{align}
The particle Green function $G_p$ describes the propagation of a particle state and
therefore is defined for $E > \mu$, $\mu$ being the chemical potential\footnote{Note that the definition of the thermodynamic limit 
(${\cal N}\to \infty$, $V\to \infty$ but ${\cal N}/V$ constant) implies $\mu({\cal N}+1)= \mu({\cal N})+ {\cal O}({\cal N}^{-1})$.},
whereas $G_h$ is defined for $E\le \mu$~\cite{fetterwalecka}.

In the limit of large momentum transfer, where the effect of collective excitation modes is expected to be negligible, 
the polarization propagator in nuclear matter reduces to
\begin{align}
\Pi({\bf q},\omega) = 2i V \int \frac{d^3p}{(2\pi)^3} \frac{dE}{2\pi}\ G({\bf p},E)G({\bf p+q}, \omega+E) \label{eq:def-pola}
\end{align}
where the discrete sum $\sum_{\bf p}$ has been replaced by $V \int d^3p/ (2\pi)^3$, 
with $V$ being the volume of the system, and the factor 2 stems from the spin sums. 
The nucleon-density response for positive excitation energies ($\omega > 0$) is then given by
\begin{align}
S({\bf q},\omega)&=-\frac{2 V}{\pi^2} \int \frac{d^3p}{(2\pi)^3} dE\  {\rm Im}G_h({\bf p},E)\nonumber\\
&\times {\rm Im}G_p({\bf p+q}, \omega+E). \label{eq:Saux}
\end{align}
The hole and particle SFs are related to the imaginary-part of the corresponding 
Green's functions through
\begin{align}
P_h({\bf p},E)=& + \frac{1}{\pi}{\rm Im}G_h({\bf p},E),\quad E \le  \mu \nonumber\\
P_p({\bf p},E)=& - \frac{1}{\pi}{\rm Im}G_p({\bf p},E), \quad E > \mu\, ,
\label{sf:greenf}
\end{align}

Introducing ${\bar P}_h({\bf p},E) = 2 V P_h({\bf p},E)/ {\cal N}$, normalized as 
\begin{equation}
\int\frac{d^3p}{(2\pi)^3}dE{\bar P}_h({\bf p},E) = 1\, ,
\end{equation}
and using Eq.(\ref{sf:greenf}), the nucleon-density response reads
\begin{align}
S({\bf q},\omega)&={\cal N}\int\frac{d^3p}{(2\pi)^3}dE{\bar P}_h({\bf p},E)\nonumber\\
&\times P_p({\bf p+q},E+\omega)\, 
\label{dens:resp}
\end{align}

When a relativistic fermion propagator is employed, its imaginary part is a matrix in the Dirac space and
contains the factor $(\slashed{p}+m)/2 e({\bf p})$, with $e({\bf p})= \sqrt{m^2+|{\bf p}|^2}$,
which can be rewritten as $[(\slashed{p}+m)/2 m] \times [m / e({\bf p})]$. The first term
enters in the matrix elements of the external current, while the second one is 
included in the definition of the nucleon-density response
\begin{align}
&S({\bf q},\omega)= {\cal N} \int \frac{d^3p}{(2\pi)^3}\ dE\ \frac{m}{e({\bf p})}\frac{m}{e({\bf p+q})}\nonumber\\
& \times {\bar P}_h({\bf p},E)P_p({\bf p+q},E+\omega)\ .
\label{dens:resp-rel}
\end{align}
The factors $m/e({\bf p})$, which reduces to one in the non-relativistic limit,
become relevant when the struck particle is relativistic. 

The GRFG SFs
\begin{align}
\label{sf:h:FG}
{\bar P}_h^{\rm GRFG}({\bf p},E)=& \frac{6\pi^2}{p_F^3}\theta(p_F-|{\bf p}|)\delta(E-e({\bf p}))\\
\label{sf:p:FG}
P_p^{\rm GRFG}({\bf p},E)=& \theta(|{\bf p}|-p_F)\delta (E-e({\bf p}))\ ,
\end{align}
yield to the scaling function of Eq.~\eqref{eq:fgrfg}.

\section{Nucleon density response and SFs in the LFG approach}
\label{sec:lda}
The LFG approach relies on the LDA, in which finite nuclei are locally treated as uniform nuclear matter of density $\rho(r)$~\cite{Nieves:2004wx, Nieves:2017lij}. Within this scheme, the density response of the nucleus is obtained integrating over its density profile 
\begin{align}
\text S^{LDA}({\bf q},\omega) &= \frac{\theta(\omega)}{4\pi^3} \int d^3r\, \int d^3 p \int_{\mu-\omega}^{\mu} dE P_h({\bf p},E)  \nonumber\\
& \times P_p({\bf p+q},E+\omega)\, ,
\label{eq:S_LDA}
\end{align}
where it is understood that both the hole and the particle SFs depend on $\rho$. Note that $\text S^{LDA}({\bf q},\omega)$ is intimately related to the imaginary part of the Lindhard function, since $-\Pi({\bf q}, \omega)/V$ turns out to be precisely the Lindhard function (particle-hole propagator) \cite{fetterwalecka} (see Eq.~\eqref{eq:def-pola}).

In the lepton-nucleus scattering analyses of Refs.\cite{Gil:1997bm, Nieves:2004wx, Nieves:2017lij}, performed using particle and hole SFs  from the semi-phenomenological model of Ref.~\cite{FernandezdeCordoba:1991wf}, the effects of collective nuclear modes were accounted for through the random phase approximation (RPA). The latter only results in modifications of the electroweak in-medium couplings, with respect to their free values, due to the presence of strongly interacting nucleons. RPA long-range correlations take into account the absorption of the gauge boson by the nucleus as a whole, instead of by an individual nucleon. Their importance decreases as the gauge boson wave-length becomes much shorter than the nuclear size. Hence, it is natural to expect that RPA effects break scaling at low momentum transfers. However these effects should become negligible in the regime of large $|{\bf q}|$ studied in this work, and  will not be included in the present calculations
\begin{figure}[]
\centering
\includegraphics[scale=0.1]{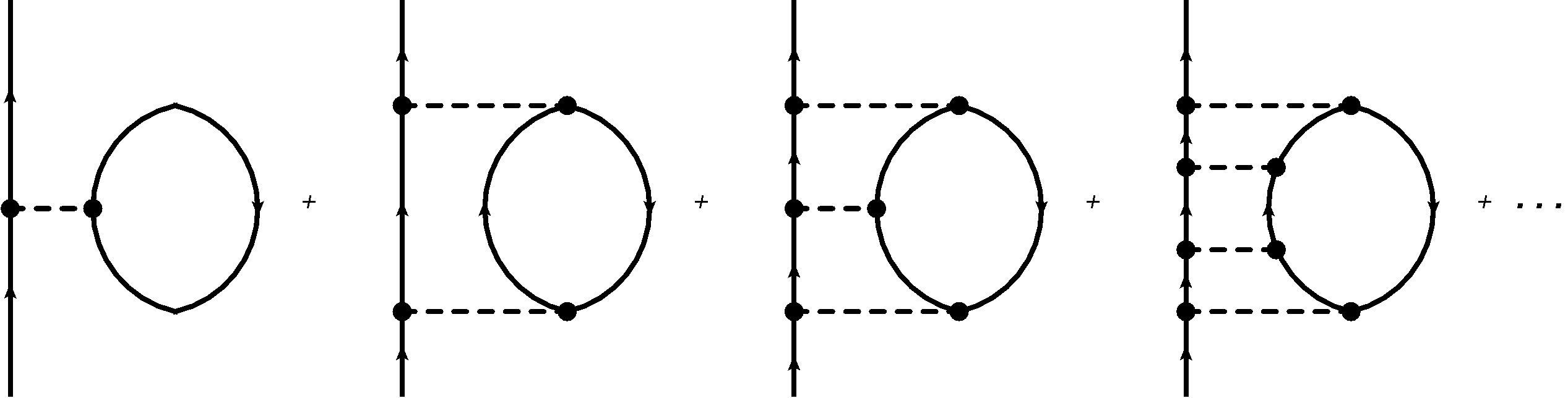}
\caption{Ladder sum of diagrams contributing to the nucleon self-energy in nuclear matter. Dashed lines represent the in-medium NN interaction.}
\label{fig:ladder_sum}
\end{figure}

The  SFs of interacting nucleons in the nuclear medium are determined by the nucleon 
self-energy $\Sigma({\bf p},E)$~\cite{Nieves:2004wx, Nieves:2017lij}
\begin{align}
&P_{p,h}({\bf p}, E) =\nonumber\\
& \mp \frac{1}{\pi} \frac{\text{Im}\Sigma({\bf p},E)}{\big(E - {\bf p}^{\,2}/2m - \text{Re}\Sigma({\bf p},E\,) \big)^2 + \text{Im}\Sigma({\bf p},E\,)^2}\ .\label{eq:SFsLDA-nonrel}
\end{align}
The chemical potential is obtained by solving the self-consistent equation  
\beq
\mu = \frac{p_F^2}{2m} + \text{Re}\Sigma(p_F, \mu)\ ,
\eeq
where the Fermi momentum of isospin-symmetric nuclear matter is given by $p_F=(3\pi\rho/2)^{1/3}$. The real part of the self-energy modifies the nucleon dispersion relation in the nuclear medium, while the imaginary part accounts for many-body decay channels. Since $\text{Im}\Sigma({\bf p},E) \ge 0 $ for $E\le\mu$, and $\text{Im}\Sigma({\bf p},E) \le 0 $ for $E > \mu$, the chemical potential can be defined as the point in which $\text{Im}\Sigma({\bf p},E)$ changes sign. 

So far we have assumed non relativistic kinematics, according to the semi-phenomenological model  for the nucleon self-energy developed in \cite{FernandezdeCordoba:1991wf}, whose  main features will be discussed in Subsec.~\ref{sec:lda_sf}. Relativistic effects can be accounted for by including the $m/e(\bf{p})$ factors in the phase space and using the relativistic expression for the nucleon energies, $e(\bf{p})$. 
In this case the hole and particle SFs read~\cite{Marco:1995vb, FernandezdeCordoba:1995pt}, 
\begin{align}
&P_{p,h}({\bf p}, E) = \nonumber \\
&\mp \frac{1}{\pi} \frac{\frac{m}{e({\bf p})}\text{Im}\Sigma({\bf p},E)}{\big(E -e({\bf p}) - \frac{m}{e({\bf p})}\text{Re}\Sigma({\bf p},E) \big)^2 + \big(\frac{m}{e(\bf{p})}\text{Im}\Sigma({\bf p},E)\big)^2},
\label{eq:lda_rel}
\end{align}
where we used the fact that in spin- and isospin-symmetric nuclear matter the self-energy operator is diagonal in the spin space. In the above equation 
$\Sigma$ stands for any matrix element $\bar u \Sigma u$, which is independent on the spin ($\bar{u}$ and $u$ are dimensionless spinors normalized to unity). 
Following the discussion below Eq.(\ref{dens:resp-rel}), the factors $m/e({\bf p})$ and $m/e({\bf p+q})$ also have to be included in the 
nucleon-density response that now reads
\begin{align}
\text S^{LDA}({\bf q},\omega) &= \frac{\theta(\omega)}{4\pi^3} \int d^3r \int d^3p\int_{\mu-\omega}^{\mu} dE \frac{m}{e({\bf p})}\nonumber\\
&\times\frac{m}{e({\bf p+q})}P_h({\bf p},E) P_p({\bf p+q},E+\omega)\, .
\label{eq:S_LDA-rel}
\end{align}

The corresponding scaling function is obtained according to Eq.~(\ref{scal:func})
\begin{equation}
f^{LDA}({\psi})= p_F\times 2\kappa\ S^{LDA}({\bf q},\omega) / \mathcal{N}
\end{equation}

\subsection{Semi-phenomenological approach to nucleon properties in nuclear matter}
\label{sec:lda_sf}
In the following, we sketch the most important features, assumptions and approximations
of the semi-phenomenological model for the self-energy developed in Ref.~\cite{FernandezdeCordoba:1991wf}, and successfully used to describe several inclusive nuclear  reactions \cite{Nieves:2004wx, Nieves:2017lij,Marco:1995vb,FernandezdeCordoba:1995pt,FernandezdeCordoba:1991yj,Gil:1997bm,Oset:1994vp,Marco:1997xb,SajjadAthar:2007bz,SajjadAthar:2009cr}. %
\begin{figure}[tbh]
\centering
\includegraphics[scale=0.1]{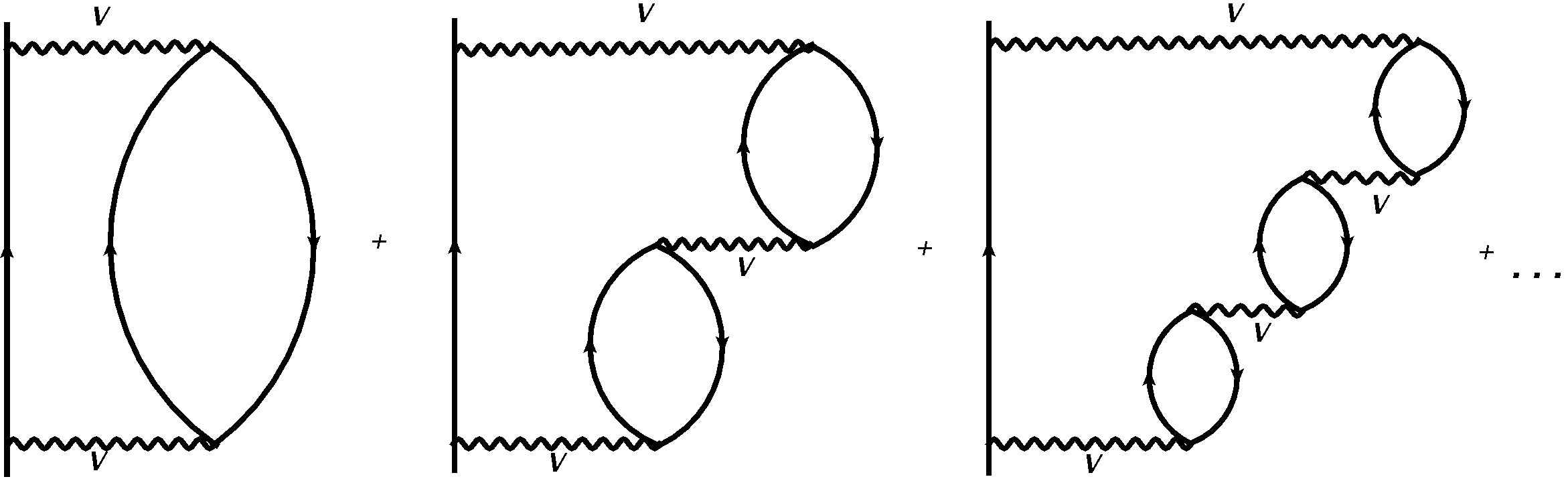}
\caption{Feynman diagrams contributing to the polarization of the NN interaction in the medium. }
\label{fig:rpa}
\end{figure}
Within this model, the non relativistic nucleon self-energy in isospin-symmetric nuclear matter is computed starting from  the low-density theorems. Short-range effects are accounted for by an in-medium effective nucleon-nucleon (NN) potential, derived from the experimental elastic NN cross section,  that in addition incorporates some medium-polarization corrections. The self-energy consists of a ladder sum of nuclear corrections generated by the series of diagrams depicted in Fig.~\ref{fig:ladder_sum}, where the dashed lines represent the effective in-medium NN potential (see Ref.~\cite{FernandezdeCordoba:1991wf} for details). Long range correlations are taken into account in the effective potential by summing up the series of diagrams shown in Fig.\ref{fig:rpa},  assuming a dominance of the transverse piece [$\tau_i \tau_j \, \sigma_i\sigma_j  (|{\bf q}|^2\delta_{ij}-q_iq_j) $] of the $ph-ph$, $ph-\Delta h$ and $\Delta h-\Delta h$ interactions~\cite{FernandezdeCordoba:1991wf}.

The imaginary-part of the self-energy, accounting for collisional broadening effects, is 
compatible with the results obtained by the more elaborate many-body calculations of Refs.~\cite{Fantoni:1983zz, Fantoni:1984zz}.
The real part of the self-energy is calculated using a dispersion relation, summing an additional Fock diagram which provides a purely real 
contribution. Only momentum-independent Hartree-type terms are missing in the model. Hence, the self-energy is determined up to an unknown momentum independent term, and it  can be used to compute in-medium nucleon properties, such as effective masses or nucleon momentum distributions, which are
found to be in good agreement with sophisticated many--body calculations~\cite{Ramos:1989hqs, Mahaux:1985zz}.

The absolute scale for the real part of the hole self-energy can be estimated from the binding energy per nucleon, $|\epsilon_A|$. 
Following Ref.~\cite{Marco:1995vb}, a phenomenological term, $C\rho$, is added to $\text{Re}\Sigma$ and
fixed against the experimental value of $|\epsilon_A|$.  With the  addition of the constant term $C\rho$, 
the chemical potential becomes 
\beq
\mu = \frac{p_F^2}{2m} + \widehat{{\cal C}}, \quad  \widehat{{\cal C}}= \text{Re}\Sigma(p_F^2/2m, p_F) + C\rho\ \label{eq:defchat}
\eeq
and the SFs read 
\begin{align}
&P_{p,h}({\bf p}, E) =\nonumber\\
& \mp \frac{1}{\pi} \frac{\text{Im}\Sigma({\bf p}, \widehat{E})}{\big(E - {\bf p}^{2}/2m - \text{Re}\Sigma({\bf p},\widehat{E})-C\rho \big)^2 + \text{Im}\Sigma({\bf p},\widehat{E}\,)^2}\label{eq:fullSLDA}
\end{align}
where $\widehat{E}\equiv E- \widehat{{\cal C}}$. The average kinetic and removal energies can be expressed in terms of the 
hole SF as~\cite{Marco:1995vb}
\begin{align}
\langle T\rangle =& \frac{4}{A}\int d^3 r\int \frac{d^3 p}{(2\pi)^3} \frac{{\bf p}^2}{2m}\int_{-\infty}^{\mu} P_h({\bf p}, E) dE\ ,\\
\langle E\rangle =& \frac{4}{A}\int d^3 r\int \frac{d^3p}{(2\pi)^3}\int_{-\infty}^{\mu} P_h({\bf p}, E) EdE\ .
\end{align}
where $A$ is the number of nucleons in the system. The binding energy per nucleon is then given by the sum  rule~\cite{Koltun:1974zz}
\beq
|\epsilon_A| = -\frac{1}{2} \bigg( \langle E\rangle +\frac{A-1}{A-2} \langle T\rangle \bigg)
\eeq
Thus for example, in carbon the parameter $C\sim 0.8$ fm$^2$, which provides around 25-30 MeV
repulsion at $\rho=0.17$ fm$^3$ and leads to  $|\epsilon_A| = 7.8$ MeV (see Table I of Ref.~\cite{Marco:1995vb}). 

Energy-dependent Dirac optical potentials for several nuclei were determined in Ref.~\cite{Cooper:1993nx} by fitting 
proton-nucleus elastic scattering data in the energy range 20-1040 MeV. In this analysis, scalar and vector complex potentials 
were employed in the Dirac equation,  and the dependences of these potentials on the kinetic energy, $t_{\rm kin}$, and radial 
coordinate, $r$, are found by fitting the scattering solutions to the measured elastic cross section, analyzing power, and spin rotation
functions.  Schr\"odinger equivalent potentials, constructed out of the scalar and vector potentials, are also given in \cite{Cooper:1993nx}. 
In Ref.~\cite{Nieves:2017lij}, the Schr\"odinger equivalent potential $^{208}$Pb central potentials obtained in Ref.~\cite{Cooper:1993nx} for $t_{\rm kin}=20$ MeV and 100 MeV
have been compared to ${\rm Re}\Sigma({\bf q}, E= {\bf q}^{2}/2m)$ from~\cite{FernandezdeCordoba:1991wf} as a function of $r$. The real 
part of the nucleon self-energy, supplemented by the kinetic-energy independent term $C\rho$, reproduced quite well the Wood-Saxon
form of the optical potentials for both values of the kinetic energy. 

It has to be noted that the results of Ref.~\cite{FernandezdeCordoba:1991wf} are not affected by the momentum-independent 
term added to the self-energy, as they only depend upon energy differences. Analogously, the nucleon-density response given 
in Eq.~\eqref{eq:S_LDA} does not depend on $C\rho$, as this term can be removed by the change of integration variable $E \to \widehat E$. 
Here, we have introduced it for the sake of comparing the LDA results with those obtained within the IA model discussed in
the next Section, where, in a first approximation, a free plane wave is used for the outgoing (ejected) nucleon. 

\section{The Impulse Approximation and the Spectral Function formalism}
\label{sec:ia}
At relatively large momentum transfer, $|{\bf q}|\gsim 500$ MeV, the IA can be safely applied under the assumption that the struck nucleon is decoupled from the spectator $(A-1)$ particles. Within this scheme~\cite{Benhar:2006wy, Benhar:2015wva}, the electromagnetic currents of Eq.(\ref{eq:RLT}) are written as a sum of one-body contributions $J_{\alpha} = \sum_i j_{\alpha}^i$ and the final nuclear state factorizes as
\beq
|f\rangle \longrightarrow |p\rangle \otimes |f\rangle_{A-1} \, .
\label{eq:fact}
\eeq
In the above equation $ |p \rangle$ is the single-nucleon state produced at the electromagnetic vertex with momentum $\mathbf{p}$, energy $e(\mathbf{p})$, and spin-isospin state $\eta_p$. The state $|f\rangle_{A-1}$ describes the residual $(A-1)$ system, its energy and recoiling momentum are fixed by energy and momentum conservation
\begin{equation}
E_{f}^{A-1}=\omega-e(\mathbf{p})+E_0\, , \quad \mathbf{P}_{f}^{A-1}=\mathbf{q}-\mathbf{p}\, .
\end{equation}
Exploiting the single-nucleon completeness relation
\begin{equation}
\sum_{k} | k\rangle \langle k| = \sum_{\eta_k}\int \frac{d^3 k}{(2\pi)^3}| \mathbf{k},\eta_k \rangle \langle \mathbf{k},\eta_k|=1\, ,
\end{equation}
and the factorization of the final state of Eq.~(\ref{eq:fact}), the matrix element of the current can be written as
\begin{equation}
\langle 0 | J_{\alpha} | f\rangle \to  \sum_{k} \langle 0| [ |k\rangle \otimes |f\rangle_{A-1}] \langle k | \sum_i j_{\alpha}^i | p \rangle\, .
\end{equation}

Substituting the last equation in Eq.~(\ref{eq:RLT}), the incoherent contribution to the response functions is given by
\begin{align}
R_{\alpha}({\bf q},\omega)&=A \sum_{p,k,k^\prime}\sum_f\, \langle k | \left(j_{\alpha}^1\right)^\dagger |p \rangle \langle p |  j_{\alpha}^1 | k^\prime \rangle\nonumber\\
&\times \langle 0 | [ |f\rangle_{A-1} \otimes |k \rangle ]  [ \,_{A-1}\langle f | \otimes \langle k^\prime| ]  | 0\rangle \nonumber\\
&\times \delta(\omega-e(\mathbf{p}) - E_{f}^{A-1} +E_0)\,\theta(|\mathbf{p}|-k_F)\, .
\end{align}

Momentum conservation in the single-nucleon vertex implies $\mathbf{k}=\mathbf{k^\prime}=\mathbf{p}-\mathbf{q}$. Charge conservation 
and the assumption that the nuclear ground state is a zero-spin state imply $\eta_k=\eta_{k^\prime}$. Therefore, using the identity
\begin{align}
\delta(\omega-e(\mathbf{p}) - E_{f}^{A-1} +E_0) &= \int dE \delta (\omega+E-e(\mathbf{p}))\nonumber\\
&\times \delta(E+E_{f}^{A-1}-E_0)\, ,
\end{align}
the response functions can be expressed as
\begin{align}
R_{\alpha}&({\bf q},\omega)=A \sum_{\eta_k\eta_p} \int \frac{d^3k}{(2\pi)^3} dE \bar{P}_h(\mathbf{k},\eta_k,E)\nonumber\\
&\times |\langle \mathbf{k},\eta_k | j_{\alpha}^1 |\mathbf{k}+\mathbf{q},\eta_p\rangle|^2 \delta(\omega+E-e(\mathbf{k+q}))\, .
\end{align}
The hole SF
\begin{align}
\bar{P}_h(\mathbf{k},\eta_k,E)&=\sum_f |\langle 0| [|\mathbf{k},\eta_k\rangle \otimes |f\rangle_{A-1}]|^2\nonumber\\
&\times\delta(E+E_{f}^{A-1}-E_0)
\label{eq:Ph}
\end{align}
gives the probability distribution of removing a nucleon with momentum ${\bf k}$ and spin-isospin $\eta_k$ from the target nucleus, 
leaving the residual $(A-1)$ system with an energy $E_0-E$.

For closed-shell nuclei and isospin-symmetric nuclear nuclear matter, the SFs of spin-up and spin-down nucleons coincide. In addition, neglecting the 
Coulomb interactions and the other (small) isospin-breaking terms, the proton and neutron SFs turn out to be identical, yielding
\begin{align}
\bar{P}_h(\mathbf{k},\eta_k,E)\simeq \frac{1}{4}\bar{P}_h(\mathbf{k},E)&=\sum_f |\langle 0| [|\mathbf{k}\rangle \otimes |f\rangle_{A-1}]|^2\nonumber\\
&\times\delta(E+E_{f}^{A-1}-E_0)
\end{align}

In order to make contact with the definition of the hole SF given in Sec.~\ref{sec:formalism}, we use the Sokhotski-Plemelj theorem~\cite{weinberg1995quantum} 
\begin{align}
\bar{P}_h(\mathbf{k},E)&=\frac{1}{\pi}\sum_f\text{Im}\langle 0| \frac{1}{E+E_{f}^{A-1}-E_0-i\epsilon} [|\mathbf{k}\rangle\nonumber\\
&\otimes |f\rangle_{A-1}][ \,_{A-1} \langle f|\otimes\langle \mathbf{k}|] |0\rangle
\end{align}
Exploiting the fact that $H|f\rangle_{A-1}=E_{f}^{A-1}|f\rangle_{A-1}$ and the completeness of the $A-1$ states, we get
\begin{align}
\bar{P}_h(\mathbf{k},E)=\frac{1}{\pi}\text{Im}\langle 0| a_{\mathbf{k}}^\dagger  \frac{1}{E+(H-E_0)-i\epsilon} a_{\mathbf{k}}|0\rangle
\end{align}
that is consistent with Eqs. (\ref{greens-fun}) and (\ref{sf:greenf}).

In the relativistic regimes, the factors  $m/e({\bf k})$ and $m/e({\bf k+q})$ have to be included to account for the implicit covariant 
normalization of the four-spinors of the initial and final nucleons in the matrix elements of the relativistic current $j_\alpha$ (see also discussion 
of Eq.~\eqref{dens:resp-rel}), hence 
\begin{align}
R_{\alpha}({\bf q},\omega)&=\frac{A}{4} \sum_{\eta_k\eta_p} \int \frac{d^3k}{(2\pi)^3} dE \bar{P}_h(\mathbf{k},E)\frac{m}{e({\bf k})}\frac{m}{e({\bf k+q})}\nonumber\\
&\times |\langle\mathbf{k}+\mathbf{q},\eta_p|  j_{\alpha}^1 | \mathbf{k},\eta_k\rangle|^2 \delta(\omega+E-e(\mathbf{k+q}))\nonumber\\
&\times\theta(|\mathbf{k+q}|-k_F)\, .
\label{eq:wmunu-IA}
\end{align}
The nucleon-density response case is recovered by $j_{\alpha}^1 \to \varrho_{\bf q}$. Carrying out the spin-isospin trace gives a factor $2$, hence
\begin{align}
S^{IA}({\bf q},\omega)&=\mathcal{N}\int \frac{d^3k}{(2\pi)^3} dE \bar{P}_h(\mathbf{k},E)\frac{m}{e({\bf k})}\frac{m}{e({\bf k+q})}\nonumber\\
&\times\delta(\omega+E-e(\mathbf{k+q}))\,\theta(|\mathbf{k+q}|-k_F)\, .
\label{eq:dens_ia}
\end{align}
Note that, within the IA the ejected nucleon is treated as a plane wave and the particle SF coincides with the one of the GRFG model given in Eq.~\eqref{sf:p:FG}. 
In analogy with Eq.~\eqref{scal:func} we can define the following scaling function
\begin{align}
f^{IA}({\psi})= p_F\times 2\kappa\ S^{IA}({\bf q},\omega) / \mathcal{N}\, .
\label{fpsi:IA}
\end{align}
The longitudinal and transverse IA scaling functions $f_{L,T}(\psi)$ can be obtained as
in Eq.~\eqref{eq:scal_func}. From Eq.~(\ref{eq:wmunu-IA}), it can be readily seen that in the IA, zeroth
kind scaling, {\em i.e.} $f^{IA}_{L}(\psi)=f^{IA}_{T}(\psi)$, only occurs if the matrix elements $\langle\mathbf{k}+\mathbf{q},\eta_p|  j_{\alpha}^1 | \mathbf{k},\eta_k\rangle$
do not depend on $k$, but only on $\bf q$ and $\omega$. Otherwise, the cancellation with the Fermi-gas 
prefactors is no longer exact. 

\subsection{Calculation of the hole SF using a correlated basis function}
\label{NMBT:sf}
The hole SF does not depend on the momentum transfer, hence it can be safely computed within non relativistic many-body theory, where 
nuclear dynamics is described by the Hamiltonian 
\beq
\label{NMBT:ham}
H=\sum_{i}\frac{{\bf p}_i^2}{2m}+\sum_{j>i} v_{ij}+ \sum_{k>j>i}V_{ijk}\ .
\eeq
In the above equation ${\bf p}_i$ is the momentum of the $i$-th nucleon, while the potentials $v_{ij}$ and $V_{ijk}$ describe two- and three-nucleon interactions,
respectively. Realistic two-body potentials are obtained from accurate fits to the available data on the deuteron and NN scattering, and reduce to the Yukawa one-pion-exchange interaction at large distances. The state-of-the-art phenomenological parametrization of $v_{ij}$, referred to as Argonne $v_{18}$ potential~\cite{Wiringa:1994wb}, is written in the form
\beq
\label{2nucl:pot}
v_{ij}= \sum_{n=1}^{18}v^n(r_{ij})O^n_{ij},
\eeq
with $r_{ij}= |{\bf r}_i - {\bf r}_j|$ and
\beq
\label{oper}
O^{n\leq 6}_{ij}= [1, ({\bm \sigma}_i\cdot {\bm \sigma}_j),S_{ij} ]\otimes [1, ({\bm \tau_i}\cdot{\bm \tau_j})] \ ,
\eeq
where ${\bm \sigma}_i$ and ${\bm \tau}_i$ are Pauli matrices acting in the spin and isospin space, respectively, and $S_{ij}$ is the tensor operator given by
\beq
S_{ij}= \frac{3}{r^2_{ij}}({\bm \sigma}_i\cdot {\bf r}_{ij})({\bm \sigma}_j \cdot{\bf r}_{ij})- ({\bm \sigma}_i\cdot {\bm \sigma}_j)\ .
\eeq
The operators corresponding to $n= 7-14$ are associated to non-static components of the NN interaction, while those corresponding to $n=15-18$ account for small violations of charge symmetry. The inclusion of $V_{ijk}$ is needed to explain the binding energies of the three-nucleon systems and nuclear matter saturation properties~\cite{Pudliner:1997ck, Pieper:2001ap}. 

In Refs.~\cite{Benhar:1989aw, Benhar:1994hw}, the nuclear overlaps, $\langle 0| [|\mathbf{k}\rangle \otimes |f\rangle_{A-1}]$, involving the ground-state and a non relativistic $1h$ and $2h1p$ states were evaluated using the CBF theory. Within this formalism, a set of correlated states (CB) is introduced
\beq
\label{corr:st}
|n\ra_{\rm CB}= \frac{{\mathcal F} |n\ra}{\la n |{\mathcal F}^\dagger {\mathcal F}|n \ra^{1/2}} \ , 
\eeq
where $|n\ra$ is an $n$ independent particle state, generic eigenstate of the free Fermi gas (FG) Hamiltonian,  and the many-body correlation operator ${\mathcal F}$ is given by
\beq
{\mathcal F}= {\mathcal S} \Big[ \prod_{j>i=1}^A F_{ij} \Big] \ .
\eeq
The form of the two-body correlation operator $F_{ij}$, reflects the complexity of the NN potential
\beq
F_{ij}= \sum_{n=1}^6 f^n(r_{ij})O^n_{ij}\ ,
\eeq
with $O^{n\leq 6}_{ij}$ given in Eq.~\eqref{oper}. The CB states are first orthogonalized (OCB)~\cite{Fantoni:1988zz} preserving, in the thermodynamical limit, the diagonal matrix elements between CB states. Then, standard perturbation theory is used to express the eigenstates of the many-body Hamiltonian of Eq.~\eqref{NMBT:ham} in terms of the OCB. Any eigenstate has a large overlap with the $n-$hole-$m-$particle OCB and hence perturbation theory in this basis is rapidly converging. 

The nuclear-matter SF can be conveniently split into two components, displaying distinctly different energy dependences~\cite{Benhar:2006wy, Benhar:2015wva, Benhar:1990zz, Benhar:1994hw}. The single-particle one, associated to one-hole ($1h$) states in $|f\rangle_{A-1}$ of Eq.~(\ref{eq:Ph}), exhibits a collection of peaks corresponding to the energies of the single-particle states belonging to the Fermi sea. 
The continuum, or correlation, component corresponds to states involving at least two-hole\textendash one-particle ($2h-1p$) contributions in $|f\rangle_{A-1}$. Its behavior as a function of $E$ is smooth and it extends to large values of removal energy and momentum~\cite{Benhar:1989aw}. It has to be noted that the correlated part would be strictly zero if nuclear correlations were not accounted for.

The carbon SF employed in this work has been computed following Ref.~\cite{Benhar:1994hw} and it is comprised of two contributions
\begin{align}
\bar{P}_h({\bf k},E)= \bar{P}^{ 1h}_h({\bf k},E) + \bar{P}_h^{\rm corr}({\bf k},E)  \label{eq:fullPh} \ .
\end{align}

The $1h$ contribution is obtained from a modified mean-field scheme
\begin{align}
\label{Pke:MF}
\bar{P}^{1h}_h({\bf k},E) = \sum_{\alpha\in \{{\rm F}\}} Z_\alpha |\phi_\alpha({\bf k})|^2 F_\alpha(E-e_\alpha) \ , 
\end{align}
where the sum includes all occupied single-particle states, labeled by the index $\alpha$, and $\phi_\alpha({\bf k})$  is the Fourier transform of the shell-model orbital with energy $e_\alpha$. Note that $|\phi_\alpha({\bf k})|^2$  yields the probability of finding a nucleon with momentum ${\bf k}$ in the state $\alpha$. The {\it spectroscopic} factor $Z_\alpha < 1$ and the function $F_\alpha(E-e_\alpha)$, describing the energy width of the state $\alpha$, account for the effects of residual interactions that are not included in the mean-field picture. In the absence of residual interactions, $Z_\alpha \to 1$  and $F_\alpha(E-e_\alpha) \to \delta_\alpha(E-e_\alpha)$. The spectroscopic factors and the widths of the $s$ and $p$ states of $^{12}$C have been taken from the analysis of  $(e,e^\prime p)$ data carried out in Refs.~\cite{Mougey:1976sc, Dutta:2003yt}.

As for the correlated part, at first CBF calculations in isospin-symmetric nuclear matter of the hole SF are carried out for several values of the density, identifying the mean-field and correlated contributions. The correlated part for finite nuclei is then obtained through an LDA procedure
\begin{align}
\label{Pke:corr}
\bar{P}^{ \rm corr}_h({\bf k},E) = \int d^3R \  \rho_A({\bf R}) \bar{P}^{\rm corr}_{h,\,NM}({\bf k},E; \rho_A({\bf R}))\,,
\end{align}
where $\rho_A(\mathbf{R})$ is the nuclear density distribution of $^{12}$C and $\bar{P}^{\rm corr}_{h\,,NM}({\bf k},E; \rho)$ is the correlation component of the SF of isospin-symmetric nuclear matter at density $\rho$. The use of the LDA to account for  $\bar{P}^{\rm corr}_h({\bf k},E)$ is based on the premise that short-range nuclear dynamics are unaffected by surface and shell effects. The energy-dependence exhibited by $\bar{P}^{\rm corr}_h({\bf k},E)$, showing a widespread background extending up to large values of both $k$ and  $E$, is completely different from that of $\bar{P}^{ 1h}_h({\bf k},E)$. For $k>p_F$, $\bar{P}^{\rm corr}_h({\bf k},E)$ coincides with $\bar{P}_h({\bf k},E)$ and its integral over the energy gives the so-called continuous part of the momentum distribution.

\subsection{Inclusion of Final State Interactions }
\label{sec:fsi}
In the kinematical region in which the interactions between the struck particle and the spectator system can not be neglected, the IA results have to be modified to include the effect of FSI. Following Ref.~\cite{Ankowski:2014yfa}, we consider the real part of the optical potential $U$ derived from the Dirac phenomenological fit of Ref.~\cite{Cooper:1993nx} to describe the propagation of the knocked-out particle in the mean-field generated by the spectator system. This potential, given as a function of the kinetic energy of the nucleon $t_{kin}(\mathbf{p})=\sqrt{{\bf p}^2+m^2}-m$, modifies the energy spectrum of the struck nucleon
\begin{equation}
\tilde{e}({\bf k+q})=e({\bf k+q})+ U\left(t_{\rm kin}({\bf k+q})\right)\, .
\end{equation}

The multiple scatterings that the struck particle undergoes during its propagation through the nuclear medium are taken into account through a convolution scheme. The IA responses are folded with the function $f_\mathbf{k+q}$, normalized as 
\beq
\int_{-\infty}^{+\infty} d\omega f_{\bf k+q}(\omega) = 1\ .
\eeq
The nucleon-density response is then given by
\begin{align}
S^{FSI}&({\bf q},\omega)={\cal N} \int \frac{d^3k}{(2\pi)^3}\, dE \int d\omega^\prime\,f_{\bf k+q}(\omega-\omega^\prime)\nonumber\\
& \times  \frac{m}{e({\bf k})}\frac{m}{e({\bf k+q})}\bar{P}_{h}({\bf k},E)\nonumber\\
&\times \delta(\omega^\prime +E-\tilde{e}({\bf k+q})) \theta(|{\bf k+q}|-p_F)\, .
\label{dens:resp:folding}
\end{align}
The scaling functions that include FSI effects are defined according to Eq.~\eqref{scal:func}
\begin{align}
f^{FSI}({\psi})= p_F\times 2\kappa\ S^{FSI}({\bf q},\omega) / \mathcal{N}\ ,
\label{fpsi:FSI}
\end{align}

Within the convolution scheme, correlations in both the hole and particle SFs are accounted for. As for the latter, comparing the above result with Eq.~\eqref{dens:resp-rel} yields 
\begin{align}
& P_p({\bf p+q},\omega+E)=\theta(|{\bf p+q}|-p_F)\nonumber\\
&\times \int d\omega^\prime\, f_{\bf p+q}(\omega-\omega^\prime)\delta(\omega^\prime +E-\tilde{e}({\bf p+q}))\, .
\label{eq:Pp_def}
\end{align}
At moderate momentum transfers, the hole and particle SFs can be consistently obtained using non relativistic many-body theory.  
However, in the kinematical region of large momentum transfer the dynamics of the struck nucleon in the final state can no longer be described using the non relativistic formalism. 
The FSI folding function is estimated employing a generalization of the Glauber theory, devised to describe high energy proton-nucleus scattering~\cite{Benhar:1991af}
\begin{align}
f_{\bf p}(\omega)=&\ \delta(\omega)\sqrt{T_{\bf p}}+\int \frac{dt}{2\pi} e^{i\omega t}\left[\bar{U}^{FSI}_{\bf p}(t)-\sqrt{T_{\bf p}}\ \right]\nonumber\\
=&\ \delta(\omega)\sqrt{T_{\bf p}}+(1-\sqrt{T_{\bf p}})F_{\bf p}(\omega)\, ,
\end{align}
where the strength of the FSI is given by the nuclear transparency $T_{\bf p}$ and the finite width function $F_{\bf p}(\omega)$. The Glauber factor $\bar{U}^{FSI}_\mathbf{p}(t)$, a detailed discussion of which can be found in Ref.~\cite{Benhar:2006wy}, is given in terms of the NN scattering amplitudes. The relation between $\sqrt{T_{\bf p}}$ and $\bar{U}^{FSI}_\mathbf{p}(t)$ can be best understood noting that~\cite{Benhar:2006wy}
\begin{align}
T_{\bf p}= \lim_{t\rightarrow\infty}P_{\bf p}(t)=\lim_{t\rightarrow\infty}|\bar{U}^{FSI}_{\bf p}(t)|^2\ ,
\end{align}
where $P_{\bf p}(t)$ is the probability that the struck nucleon does not undergo re-scattering processes during a time $t$ after the electromagnetic interaction.
In absence of FSI $\bar{U}^{FSI}_{\bf p}(t)=1$, implying in turn $ T_{\bf p}=1$ and $f_{\bf p}(\omega) \to  \delta(\omega)$.

In Ref.~\cite{Ankowski:2014yfa} the convolution scheme was further approximated, assuming that for large momentum transfer $t_{kin}(|\mathbf{k + q}|)\simeq t_{kin}(|\mathbf{q}|)$. As a consequence, the real part of the optical potential only produces a shift of the response to lower energy transfer. In this work, we retain the full dependence on $|\mathbf{p}|=|\mathbf{k + q}|$, which brings about a Jacobian when solving the angular integral of the initial momentum of the nucleon. This Jacobian, not negligible in the kinematical regime where FSI are important, quenches the quasi-elastic peak of the response, enhancing its tails. 
 
 In order to make contact with the LFG formalism of Sec.~\ref{sec:lda}, we rewrite the particle SF as
\begin{align}
P_p(\mathbf{p},E)&=\theta(|{\bf p}|-p_F)[\sqrt{T_{\bf p}}\delta(E-\tilde{e}(\mathbf{p}))\nonumber\\
&+(1-\sqrt{T_{\bf p}})F_{\bf p}(E-\tilde{e}(\mathbf{p}))]\, .
\label{eq:sfsi}
\end{align}
In the simple case of a zero-range NN interaction and neglecting correlation effects in the eikonal factor~\cite{Benhar:2013dq}
\begin{align}
F_{\bf p}(E-\tilde{e}(\mathbf{p}))=&-\frac{1}{\pi}\frac{{\rm Im}V({\bf p})}{(E-\tilde{e}(\mathbf{p}))^2+{\rm Im}V({\bf p})^2}\, ,
\end{align}
where 
\begin{align}
{\rm Im}V({\bf p})=-\frac{1}{2} \rho v_p \sigma_p\, .
\end{align}
In the above equation, $v_p=|{\bf p}|/m$ is the velocity of the struck particle, which in the eikonal approximation is assumed to be constant, $\rho$ is the average nuclear density, and $\sigma_p$ is the total NN cross section.

Under these assumptions, Eq.~\eqref{eq:sfsi} can be rewritten as 
\begin{align}
P_p(\mathbf{p},E)&\simeq \theta(|{\bf p}|-p_F)\Big[-\frac{1}{\pi} \frac{{\rm Im}V({\bf p})}{(E-\tilde{e}(\mathbf{p}))^2+{\rm Im}V({\bf p})^2} \nonumber\\
&+ \delta P_p^{FSI}\Big]\, ,
\label{part:sf:approx}
\end{align}
where
\begin{align}
\delta P_p^{FSI}&=\sqrt{T_{\bf q}} \Big[ \delta(E-\tilde{e}(\mathbf{p}))\nonumber\\
&+\frac{1}{\pi} \frac{{\rm Im}V({\bf p})}{(E-\tilde{e}(\mathbf{p}))^2+{\rm Im}V({\bf p})^2}\Big]\, .
\end{align}
The term $\delta P_p^{FSI}$ is expected to be small in large nuclei since $T_{\bf p}=0$ in infinite nuclear matter. In addition, it vanishes for ${\rm Im}V\to 0$, as in this limit the Lorentzian distribution 
cancels the $\delta$-function. Neglecting $\delta P_p^{FSI}$, the expression reported in Eq.~\eqref{part:sf:approx} is reminiscent of the definition of the SF in terms of the nucleon self-energy given in Eq.~\eqref{eq:lda_rel}. Therefore, 
the approaches discussed in Secs.~\ref{sec:lda} and \ref{sec:ia} can be  approximately connected through the following identifications 
\begin{align}
\label{eq:iden-ima}
&\theta(|{\bf p}|-p_F){\rm Im}V({\bf p})\to   \frac{m}{e({\bf p})} \text{Im}\Sigma({\bf p},{\widehat E})\, \Big |_{\rm avg}, \quad E > \mu\\
&U\left(t_{kin}({\bf p})\right)  \to \frac{m}{e({\bf p})} \text{Re}\Sigma({\bf p},{\widehat E})+ C\rho \, \Big|_{\rm avg}, \quad E > \mu
\end{align}
for some average density. The step function in Eq.~\eqref{eq:iden-ima}, which accounts for Pauli-blocking effects as in the FG model, implies that the particle SF vanishes when $|{\bf p+q}| < p_F$. We should stress 
that the LDA approach employs a dynamical particle self-energy that separately depends 
on the energy and momentum.

\section{Results}\label{results}
\begin{figure}[]
\centering
\includegraphics[scale=0.7]{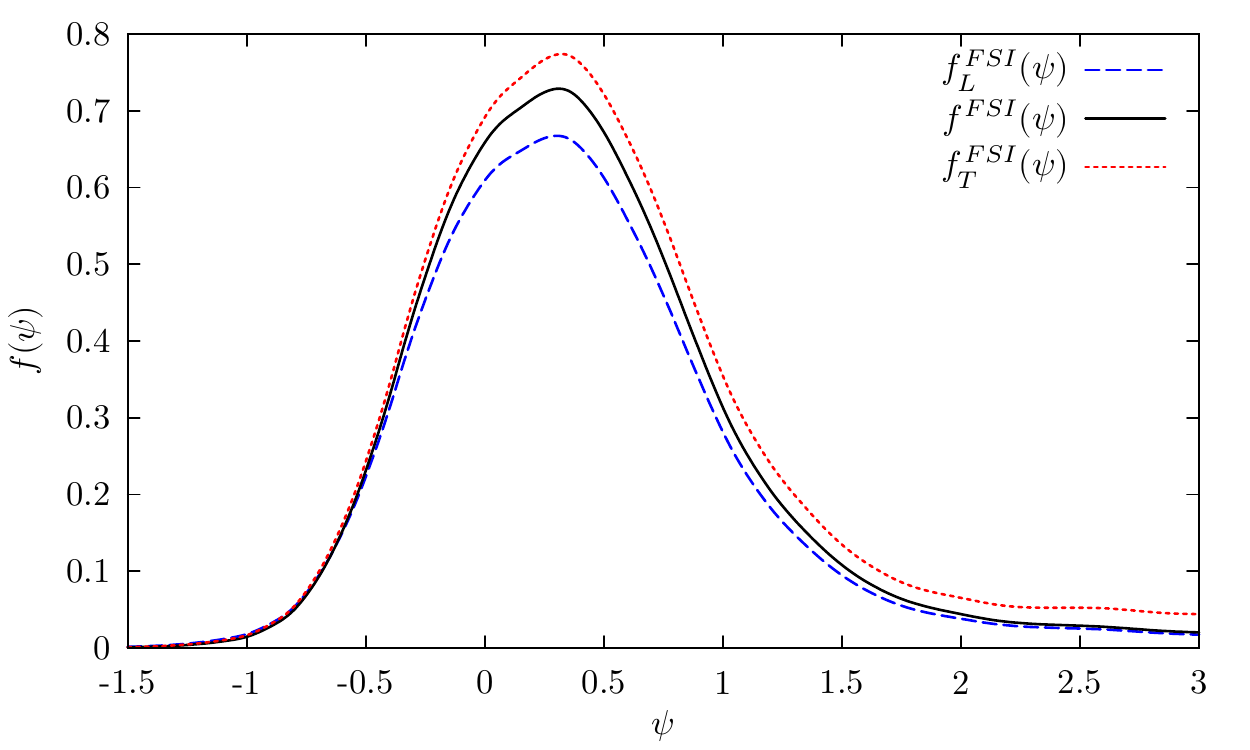}
\caption{Transverse (red dotted), longitudinal (blue dashed) and nucleon-density (black-solid) scaling functions of $^{12}$C at $|{\bf q}|=1.0$ GeV obtained from the CBF SF approach including FSI corrections.}
\label{fl_ft_f_SF}
\end{figure}

\begin{figure}[]
\centering
\includegraphics[scale=0.7]{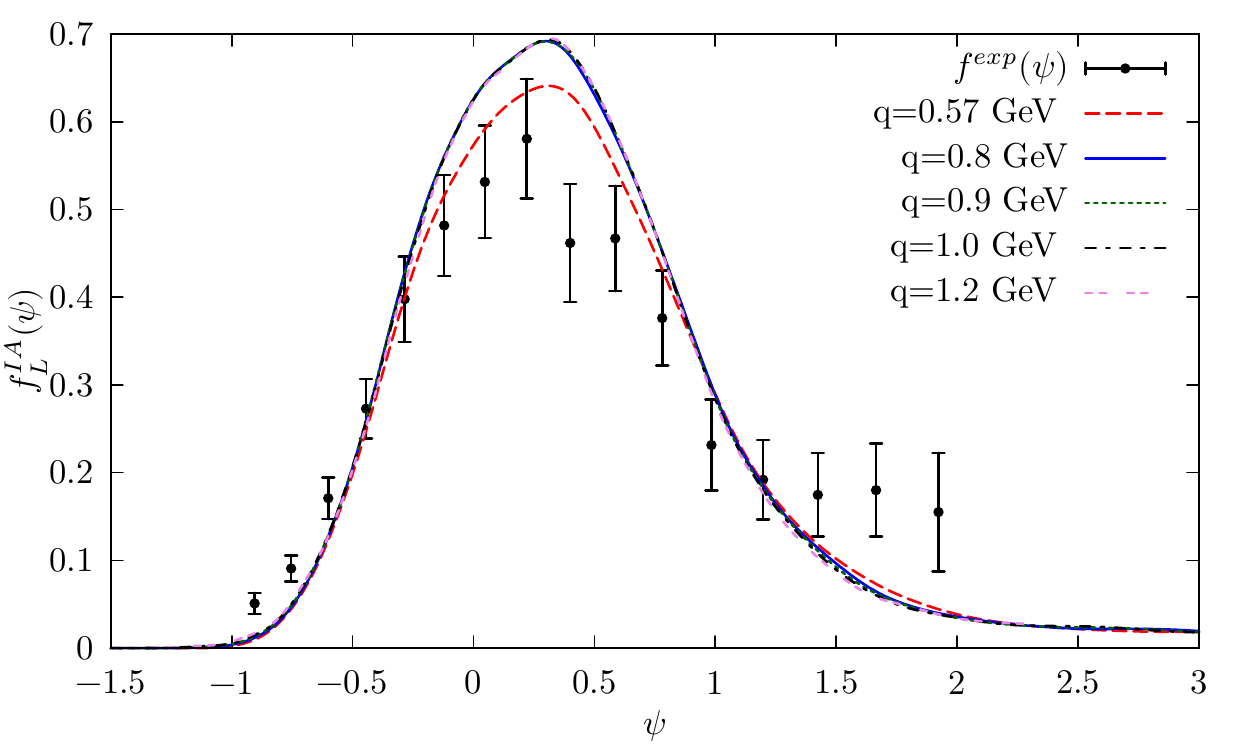}
\includegraphics[scale=0.7]{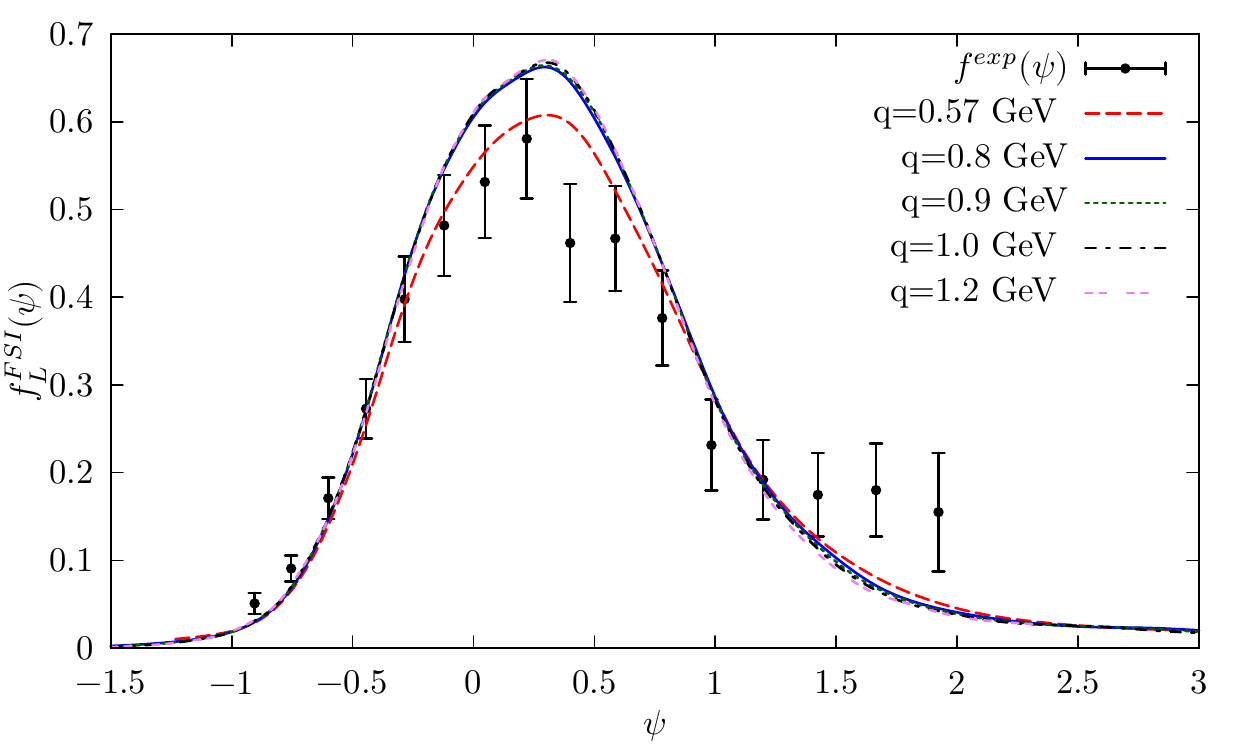}
\caption{ Longitudinal scaling functions in $^{12}$C computed using the hole SF (Eq.~\eqref{eq:fullPh}) of Ref.~\cite{Benhar:1994hw} for $|{\bf q}|=0.57,\ 0.8,\ 0.9,\ 1.0,\ {\rm and}\ 1.2$ GeV. Results obtained within the IA scheme are shown in the  upper panel, while those including FSI effects are displayed in the bottom one.  The standard definition of the longitudinal prefactor given in Eq.~(30) of Ref.~\cite{Rocco:2017hmh} has been used to get both the theoretical curves and the experimental points obtained from the $|{\bf q}|=0.57$ GeV  data of Ref.~\cite{Barreau:1983ht}}.
\label{fl_SF}
\end{figure}

In this section we present the $^{12}$C electromagnetic scaling functions obtained using the SF approaches outlined in Secs.~\ref{sec:lda} and ~\ref{sec:ia}. When defining the scaling variable $\psi$, we used $p_F=225$ MeV, accordingly to the analysis of electron-scattering data of Ref.~\cite{Maieron:2001it}. In the following we denote with ``FSI'' the results of the CBF hole SF supplemented by the convolution scheme and with ``IA'' those in which FSI are neglected, as in Eq.~\eqref{eq:dens_ia}. With ``LDA'' we indicate the semi-phenomenological approach of Sec.~\ref{sec:lda} consistently adopted for both the hole and particle SFs.  When a relativistic free nucleon in the final state (delta distribution for the particle SF) and a fully dressed hole are considered, the curves are labeled as ``IA LDA''.

In Fig.~\ref{fl_ft_f_SF}, the transverse, longitudinal and nucleon-density scaling functions obtained using the CBF SF are compared. In all cases 
FSI effects are included. Despite only one-body current contributions are considered, an enhancement in the transverse channel (red dotted curve) with respect to the longitudinal one (blue dashed curve) is apparent. The nucleon-density scaling function (solid black curve) lies between the transverse and the longitudinal ones, corroborating this choice of the scaling function. Our analysis suggests that the differences between the three curves have to be ascribed to the use of the GRFG model prefactors in the scaling functions.

FSI effects in the IA scheme can be appreciated from Fig.~\ref{fl_SF}. The IA and FSI longitudinal scaling functions at $|{\bf q}|=0.57,\ 0.8,\ 0.9,\ 1.0,\ {\rm and}\ 1.2$ GeV, obtained within the CBF SF approach using the hole SF of Ref.~\cite{Benhar:1994hw}, are displayed in the upper and bottom panels, respectively. FSI do not play a major role, leading to very small modifications of the IA results except for $|{\bf q}|=0.57$ GeV, where they improve the agreement with experimental data. Our findings are at variance with those of Ref.~\cite{Caballero:2007tz}, where the violation of zeroth-kind scaling are ascribed to relativistic effects in the FSI. The asymmetric shape of the theoretical scaling functions, mildly affected by the inclusion of FSI, is clearly visible, although less pronounced than in the data. 

\begin{figure}[!h]
\centering
\includegraphics[scale=0.7]{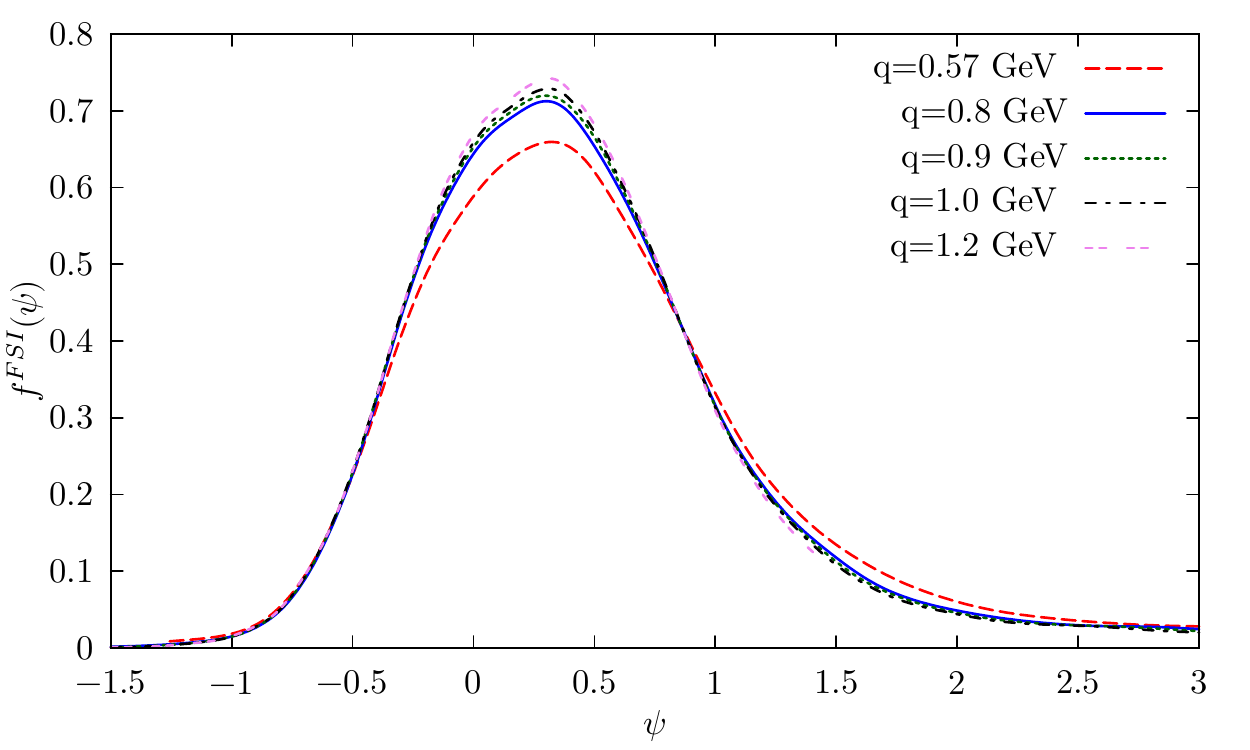}\\
\includegraphics[scale=0.7]{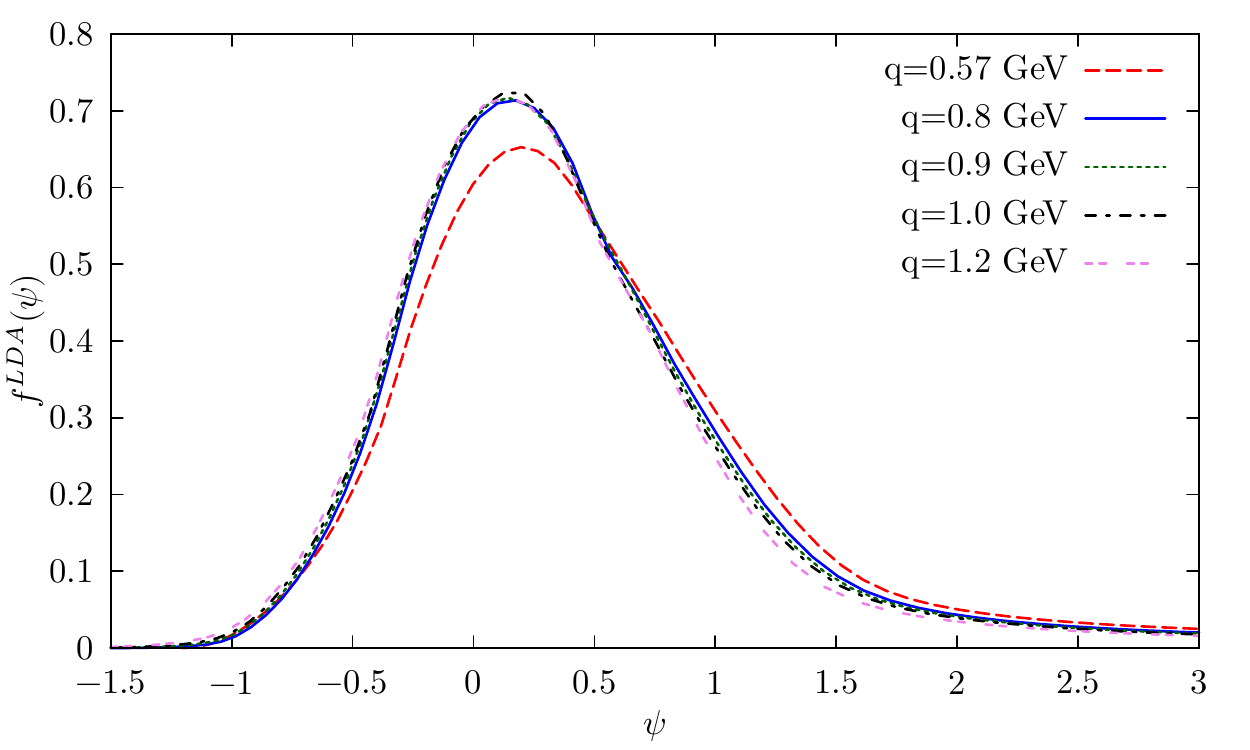}\\
\caption{Nucleon-density scaling functions for $^{12}$C computed for $|{\bf q}|=0.57,\ 0.8,\ 0.9,\ 1.0,\ {\rm and}\ 1.2$ GeV. In the upper (lower) panel, results obtained using CBF (relativized LDA) SFs are shown}
\label{fpsi_SF}
\end{figure}

\begin{figure}[]
\centering
\includegraphics[scale=0.7]{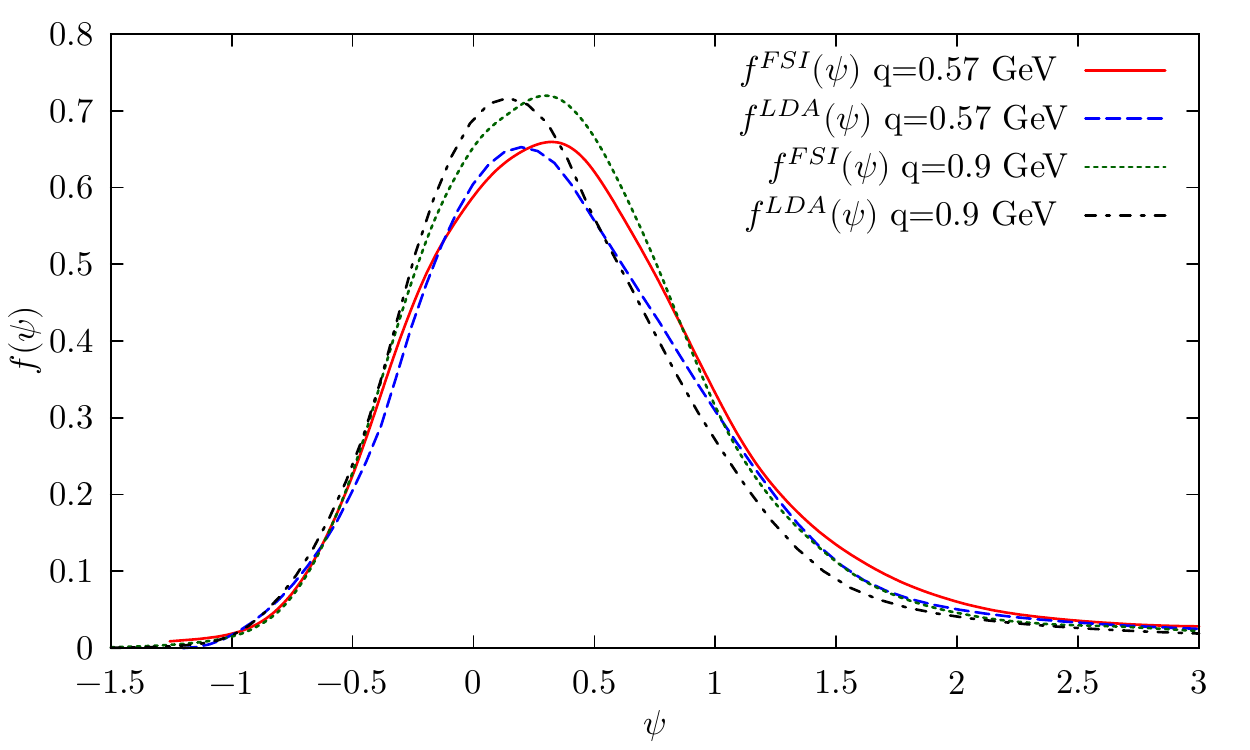}
\includegraphics[scale=0.7]{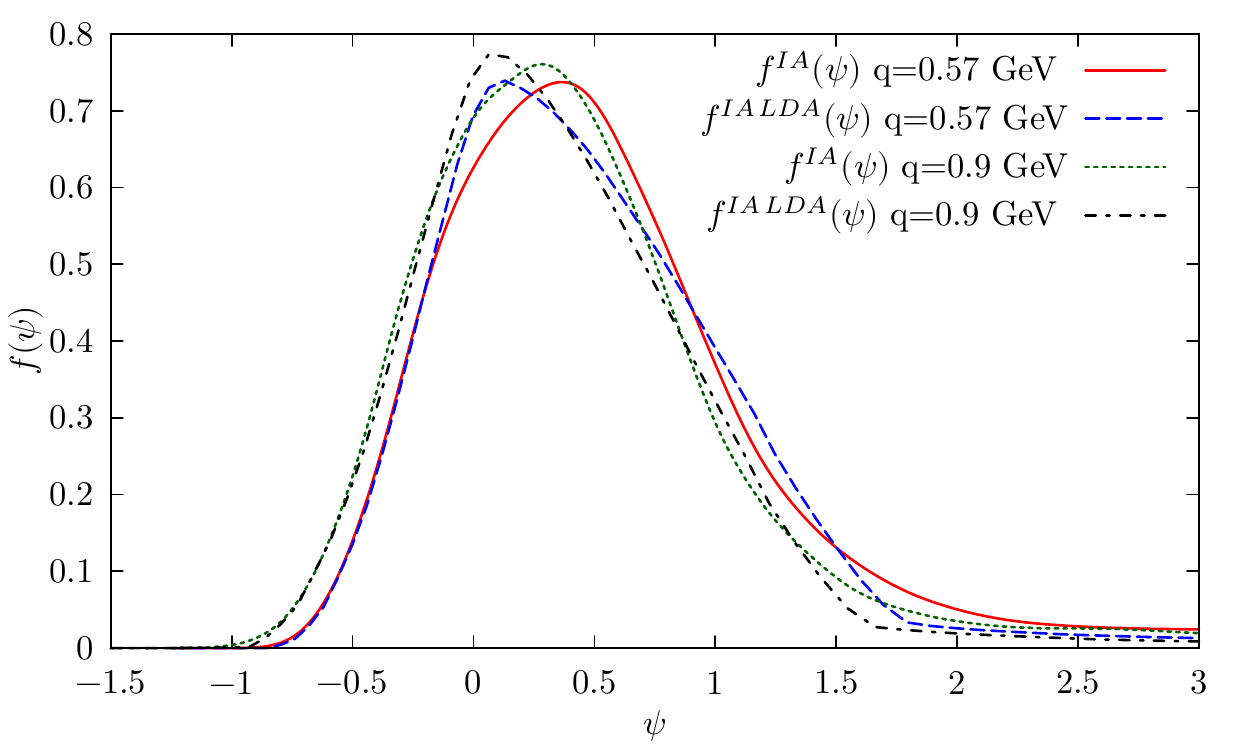}
\caption{Scaling functions for $^{12}$C computed for $|{\bf q}|=0.57,\ {\rm and}\ 0.9$ GeV. In the upper (lower) panel results are obtained within the FSI(IA) and LDA (IA LDA) approaches.}
\label{IA_LDA_12C}
\end{figure}

In Fig.~\ref{fpsi_SF} we compare the nucleon-density scaling functions obtained using the relativized LDA approach against those of the CBF SF, for the same momentum transfer values of Fig.~\ref{fl_SF}, including FSI in the two schemes. Both approaches provide asymmetric scaling functions that satisfy scaling of the first kind. The comparison between LDA and CBF predictions can be better appreciated in Fig.~\ref{IA_LDA_12C}, where results for $|{\bf q}|=0.57$  and 0.9 GeV are highlighted. In the upper panel, FSI and LDA results nicely agree for both momentum transfers. In the lower panel, we show that the consistency between the two approaches is preserved also in the IA frame, provided the $C\rho$ term is included in the real part of the LDA self-energy. Comparing the upper and lower panels, we find appreciable FSI effects only for $|{\bf q}|=0.57$ GeV. Their inclusion leads to a shift of the peak position towards smaller values of $\psi$ and to a redistribution of the strength, 
which enhances the asymmetry of the nucleon-density scaling functions. 
The differences in the position of the quasi-elastic peak -- the CBF curves are shifted towards larger excitation energies compared to those of the LDA SF -- have to be ascribed to the more accurate description of the structure of $^{12}$C provided by the CBF SF. This is encoded in the mean-field contribution $\bar{P}^{1h}_h({\bf p},E)$, extracted from $(e,e^\prime p)$ experiments, and cannot be encompassed by the LDA approach of Sec.~\ref{sec:lda}. It is also remarkable that the LDA model of Sec.~\ref{sec:lda} leads to tails of the scaling functions comparable to those arising in the CBF formalism. In the latter case, these tails are mostly provided by the correlation contribution $\bar{P}_h^{\rm corr}({\bf p},E)$ of the hole SF, and hence they are quite sensitive to short-range correlations. In the LDA approach these correlations are incorporated in the in-medium NN potential obtained from the experimental elastic NN scattering cross section, modified to include some medium polarization corrections.  

\section{Analysis} 
\label{sec:toy}
The origin of the scaling exhibited by the nuclear responses has a simple and exact formulation
within the GRFG model, which, however, largely fails to reproduce experimental data.    
Understanding the scaling features of nuclear responses becomes challenging when the nucleus is treated as
a fully-interacting many-body system. 

In order to avoid the complications arising when GRFG model prefactors are used to remove single-nucleon dynamics, we
will focus on the nucleon-density scaling function, defined in Eq.(\ref{scal:func}). To address the dynamical origin of first-kind scaling,
we will consider a simplified description of the nucleus, yet retaining the key aspects of the many-body problem. For simplicity, our analysis 
is limited to non relativistic kinematics. Hence, in the following we will use the non relativistic scaling variable~\cite{Rocco:2017hmh}
\begin{equation}
\psi^{nr} =\frac{1}{p_F} \left(\frac{m\omega}{|{\bf q}|}-\frac{|{\bf q}|}2\right)\, .
\end{equation}
A generalization to the relativistic case does not involve conceptual difficulties.

\subsection{PWIA model}
Within the IA, the non relativistic nucleon-density scaling function is defined as 
\begin{align}
f^{IA}({\bf q}, \omega)=&  2\kappa p_F \int \frac{d^3p}{(2\pi)^3} \ dE  \bar{P}_h({\bf p},E)\theta (|{\bf p+q}|-p_F)\nonumber\\
&\times \delta\left(\omega +E -e({\bf p+q})\right) \ , \label{eq:fpwia}
\end{align}
where $e({\bf p})$ is the non relativistic energy spectrum of the initial nucleon with momentum ${\bf p}$. 

The above expression can be further simplified within the plane wave impulse approximation (PWIA), which amounts to neglect 
information on the target removal energy distribution. The hole SF is written in the approximate form 
\begin{equation}
 \bar{P}_h({\bf p},E) \simeq \bar{n}({\bf p}) \delta\left(E-e({\bf p})\right)\, ,
\end{equation}
where the momentum distribution is defined as
\begin{equation}
\bar{n}({\bf p})=\int dE  \bar{P}_h({\bf p},E)\quad,\quad \int \frac{d^3p}{(2 \pi)^3} \bar{n}({\bf p})=1\, .\label{eq:onebody-appr}
\end{equation}
We will use a state-of-the-art momentum distribution computed within variational Monte Carlo in
Ref.~\cite{Wiringa:2013ala}. 

Within the PWIA, the nucleon-density scaling function reads
\begin{align}
f^{PWIA}({\bf q}, \omega) &=2\kappa p_F \int \frac{d^3p}{(2\pi)^3}\ \overline{n}({\bf p})\theta (|{\bf p+q}|-p_F)\nonumber\\
&\times\delta\left(\omega + e({\bf p})- e({\bf p+q})\right)\, .
\label{eq:f_pwia}
\end{align}

To better elucidate the emergence of first-kind scaling and the asymmetry of the scaling function,  
we consider three different scenarios with increasing sophistication for the description of the energy spectrum.

Let us first assume a free energy spectrum for both the hole and particle states in the energy-conserving $\delta$ function
\begin{align}
\delta(\omega +e({\bf p})-e(|{\bf p+q}|))=\delta\Big(\omega- \frac{|{\bf q}|^2}{2m}-\frac{|{\bf p}||{\bf q}|\cos\theta}{m}\Big)\ ,
\label{eq:delta-free-approx}
\end{align}
where $\theta$ is the angle between ${\bf p}$ and ${\bf q}$. 
The integration over $\cos\theta$ can be performed using the $\delta$-function, which gives rise to a Jacobian 
\begin{align}
\mathcal{J}= \frac{m}{|{\bf p}||{\bf q}|} = \frac{1}{2|{\bf p}|\kappa}\, .
\label{jac:free}
\end{align}
The fact that $|\cos\theta|\leq 1$ provides a lower bound to the momentum of the hole
\begin{equation}
|{\bf p}| \ge p_F \ |\psi^{nr}|\, .
\label{eq:lowerbound-free}
\end{equation}
An additional constraint comes from the step function $\theta (|{\bf p+q}|-p_F)=\theta(e({\bf p})+\omega-p_F^2/2m)$, yielding
\begin{equation}
|{\bf p}|^2 \ge p_F^2-2m\omega\,.
\label{eq:lb_2}
\end{equation}
The latter constraint is always satisfied for sufficiently large values of $\omega$, in which case the integration range of $|{\bf p}|$
is limited by Eq.~\eqref{eq:lowerbound-free} only. For low momentum and energy transfers, the lower limit is instead
the one of Eq.~\eqref{eq:lb_2} leading to violations of first-kind scaling, unless a piecewise definition of  $|\psi^{nr}|$
is adopted~\cite{Alberico:1988bv}.

Since the factor $\kappa$ that appears in Eq.~\eqref{eq:f_pwia} simplifies with the Jacobian, the result of the
integration only depends upon the lower integration limit, $p_F |\psi^{nr}|$, and thus it is easily found that $f^{PWIA}$ is a symmetric 
function of $\psi^{nr}$, as it only depends on the modulus of this variable. 

\begin{figure}[h!]
\includegraphics[scale=0.675] {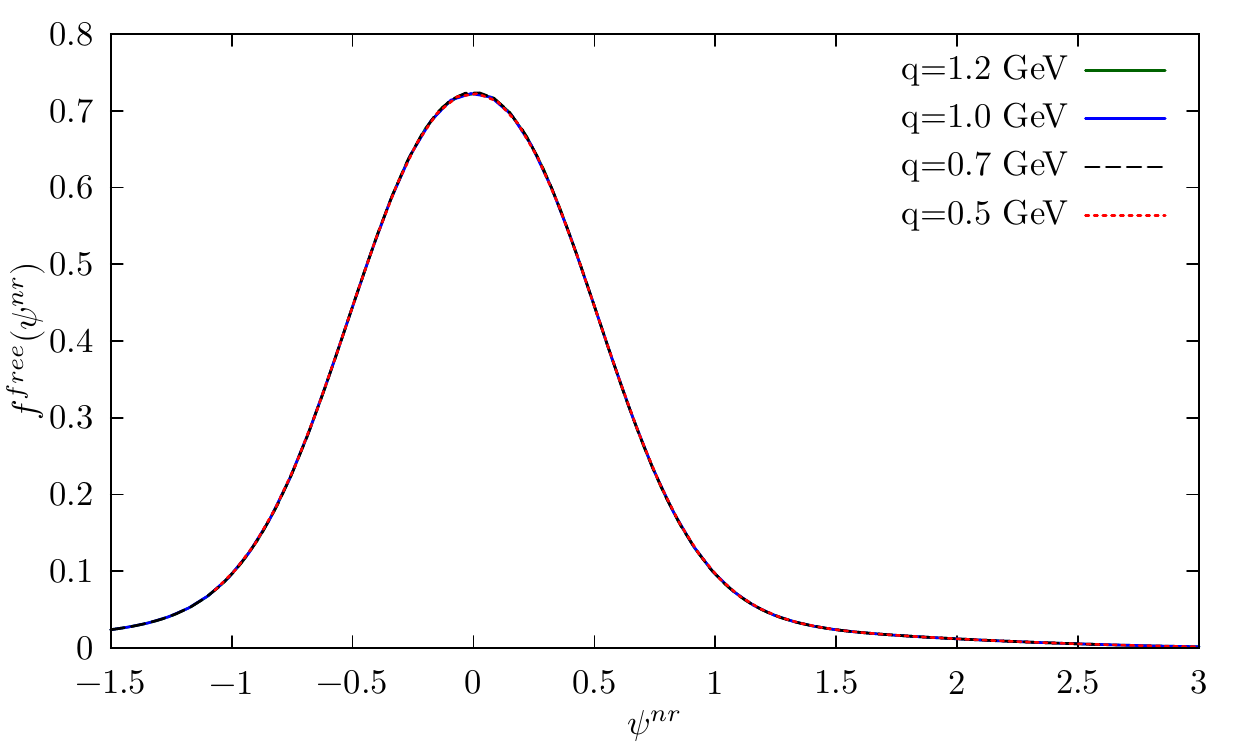} 
\caption{Non-relativistic PWIA scaling responses, using the momentum distribution of $^{12}$C derived in Ref.~\cite{Wiringa:2013ala} for $|{\bf q}|= 0.5,\ 0.7,\ 1$ and 1.2 GeV. The Fermi momentum has been fixed to $p_F=225$ MeV.}
\label{scal_NM_free}
\end{figure}

Figure \ref{scal_NM_free} shows the PWIA nucleon-density scaling functions of $^{12}$C, using the energy-conserving $\delta$ function of Eq.~\eqref{eq:delta-free-approx}, for different momentum transfers. Scaling is perfectly satisfied: the curves are peaked around  $\psi^{nr}=0$ and do not show any asymmetry, as expected from the above discussion. The only difference with the GRFG case is that the scaling function extends to values of $|\psi^{nr}|$ larger than 1. This is due to the fact that $\bar{n}(\mathbf{p})$ does not vanish above $p_F$.

\begin{figure}[h!]
\includegraphics[scale=0.7] {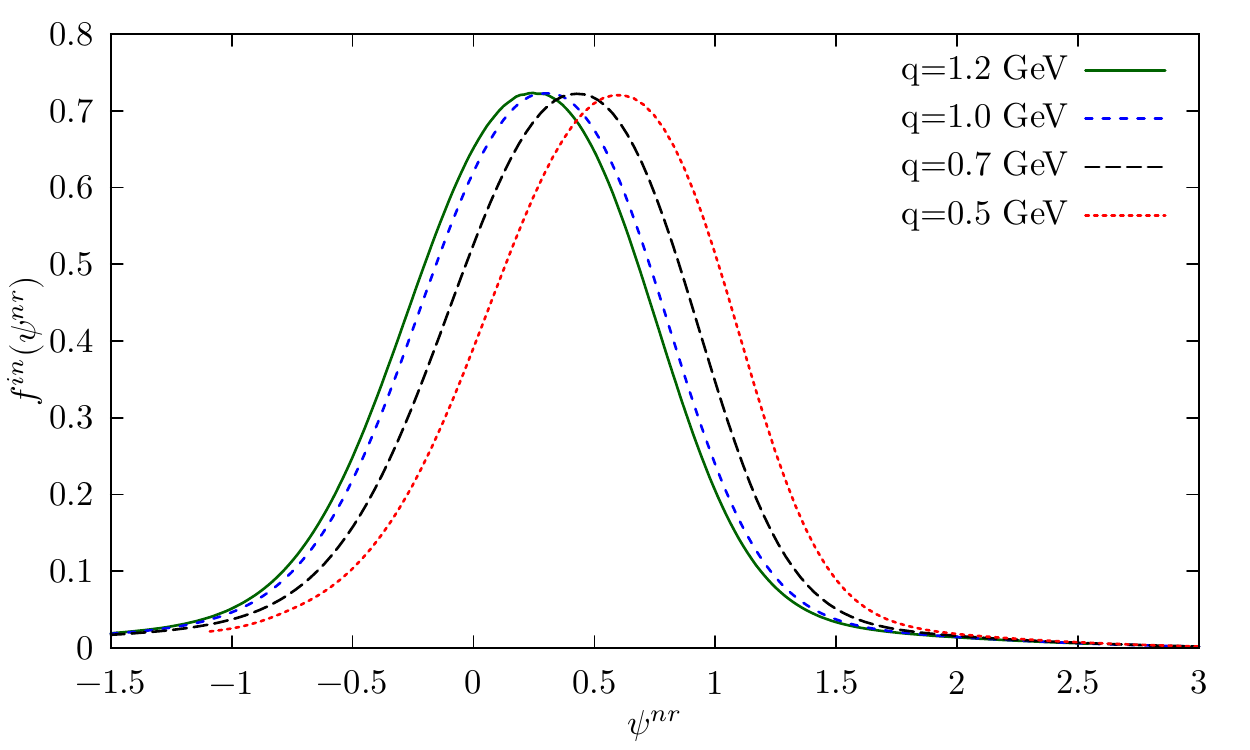} 
\caption{Non relativistic scaling responses obtained within PWIA (Eq.~\eqref{eq:f_pwia}) as a function of $\psi^{nr}$ for $|{\bf q}|= 0.5,\ 0.7,$ and 1 GeV. 
The momentum distribution of $^{12}$C derived in Ref.~\cite{Wiringa:2013ala} has been used,  and the energy of the hole state 
has been extracted from the calculations of the nuclear matter energy spectrum of Ref.~\cite{Wiringa:1988jt} and implemented in the energy conservation (see Eq.~\eqref{scal:in:corr}). 
The Fermi momentum, $p_F$, has been fixed to 225 MeV.}
\label{scal_NM_in}
\end{figure}

As a second step, we treat the hole as a bound state using the energy spectrum of nuclear matter at saturation density of Ref.~\cite{Wiringa:1988jt} (see also the recent work of Ref.~\cite{Benhar:2017oli}). In this case the energy conserving $\delta$-function is given by
\begin{align}
\delta\left(\omega + \mathcal{U}({\bf p})- \frac{|{\bf q}|^2}{2m}-\frac{|{\bf p}||{\bf q}|\cos\theta}{m}\right)\ .
\label{scal:in:corr}
\end{align}
where the single-particle potential $\mathcal{U}({\bf p})<0$ has been added to $e({\bf p})$. Modifying the hole energy spectrum does not change the Jacobian of Eq.~\eqref{jac:free}. However, the lower bound of Eq.~\eqref{eq:lowerbound-free} now reads 
\begin{equation}
|{\bf p}| \ge \left | p_F \ \psi^{nr} + m\frac{\mathcal{U}({\bf p})}{|\mathbf{q}|}\right |\ . 
\end{equation}
The term $\mathcal{U}({\bf p})/|\mathbf{q}|$ introduces further dependences on $|{\bf q}|$ and leads to violations of first-kind scaling. These violations are apparent in the results displayed in Fig.~\ref{scal_NM_in}, where the $|{\bf p}|$-dependent term in the energy-conserving $\delta$-function leads to a shift of the different curves. The peaks move to higher excitation energies, as expected for an attractive average hole potential. 
For $|{\bf q}|$=1.0, 1.2 GeV, the curves peak approximately at $\psi^{nr}=0 $ and the result found in the free energy case is recovered to a very large extent. This can be easily understood, since the average  $\mathcal{U}({\bf p})^{\rm avg}/|\mathbf{q}|$ correction becomes small for large values of the momentum transfer. The shape of the scaling 
functions, which is still symmetric around $\psi^{nr}=0$, is almost unaffected by the single-particle potential.

\begin{figure}[h!]
\includegraphics[scale=0.7] {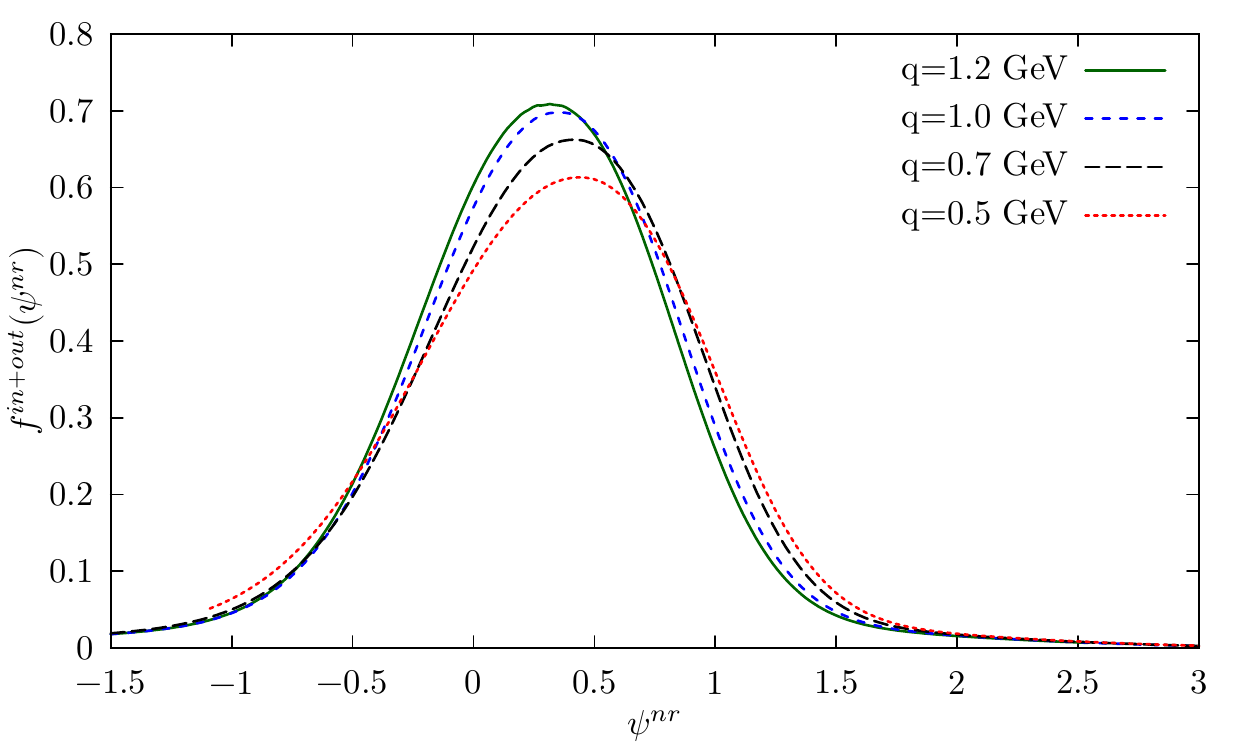} 
\caption{Same as in Fig.\ref{scal_NM_in},  but nuclear potentials have been used to determine both the hole and particle state energies (see Eq.~\eqref{scal:inout:corr}).}
\label{scal_NM_tot}
\end{figure}

Finally, we consistently include a single-particle potential in the hole and particle energy spectra. The energy conserving $\delta$-function reads
\begin{align} 
\delta\Big(\omega- \frac{|{\bf q}|^2}{2m}-\frac{|{\bf p}||{\bf q}|\cos\theta}{m} +\mathcal{U}(|{\bf p}|)-\mathcal{U}(|{\bf p+q}|)\Big)\ .
\label{scal:inout:corr}
\end{align}
The non trivial dependence on $\cos\theta$ hidden in $\mathcal{U}(|{\bf p+q}|)$ prevents, in general, from analytically solving the integral. To circumvent this problem, we performed a numerical integration, treating the $\delta$-function as the limit of a Gaussian. This allows us to properly evaluate the Jacobian, which differs from the one reported in Eq.~\eqref{jac:free}. 
This introduces a first source of  scaling violations, as the $\kappa$ factor of Eq.~\eqref{eq:f_pwia} does not exactly cancel with the Jacobian. Nevertheless, the cancellation is still partially produced and becomes exact in the $|{\bf q}| \gg |{\bf p}|$ limit. Fig.~\ref{scal_NM_tot} displays the scaling functions computed using Eq.~\eqref{scal:inout:corr} for the energy-conserving $\delta$ function for the same kinematical setups as in Figs.~\ref{scal_NM_free} and \ref{scal_NM_in}.  The curves are still shifted\footnote{The resulting breaking scaling pattern can be understood taking into account that 
\begin{equation}
 \mathcal{U}(|{\bf p}|)-\mathcal{U}(|{\bf p+q}|) <0
\end{equation}
and that in the large momentum transfer, this difference becomes independent of $\cos\theta$, and has little influence in the lower limit of the 
$|{\bf p}|-$integration.} compared to the free case, although the position of the peaks is closer to $\psi^{nr}=0$ than in Fig.~\ref{scal_NM_in}. This indicates a partial cancellation of single-particle potentials in the hole and particle spectra, as discussed in Ref.~\cite{Nieves:2017lij}. As alluded to earlier, the new Jacobian  introduces a residual dependence on $|{\bf q}|$, specifically in the magnitude of the scaling functions. First kind scaling is almost recovered for $|{\bf q}| \ge 1$ GeV, although scaling violations are already small 
for $|{\bf q}| = 0.7$ GeV.  As in the other cases, scaling functions exhibit only a small asymmetry.
 
Up to now, we have neglected the imaginary part of the in-medium potentials. As discussed in Refs.~\cite{FernandezdeCordoba:1991wf, Nieves:2017lij}, effects on the ejected-nucleon are expected to be larger than in the hole state. The corrections induced by the imaginary part of the optical potential on the particle states can be estimated, following the approach detailed in Subsec.~\ref{sec:fsi}, by convoluting  the PWIA scaling function as in Eq.~\eqref{dens:resp:folding}. Since Eq.~\eqref{scal:inout:corr} consistently includes the single-particle potential, both in the hole and particle energy spectra, the real part of the potential does not have to be included in the argument of the folding function. Analogously to the discussion in Fig.~\ref{fl_SF}, the corrections are very small and have little effects on the discussion about the origin of the scaling. Moreover, these FSI corrections do not induce any appreciable asymmetry in the scaling functions.

\subsection{Beyond PWIA}
\begin{figure}[]
\includegraphics[scale=0.675] {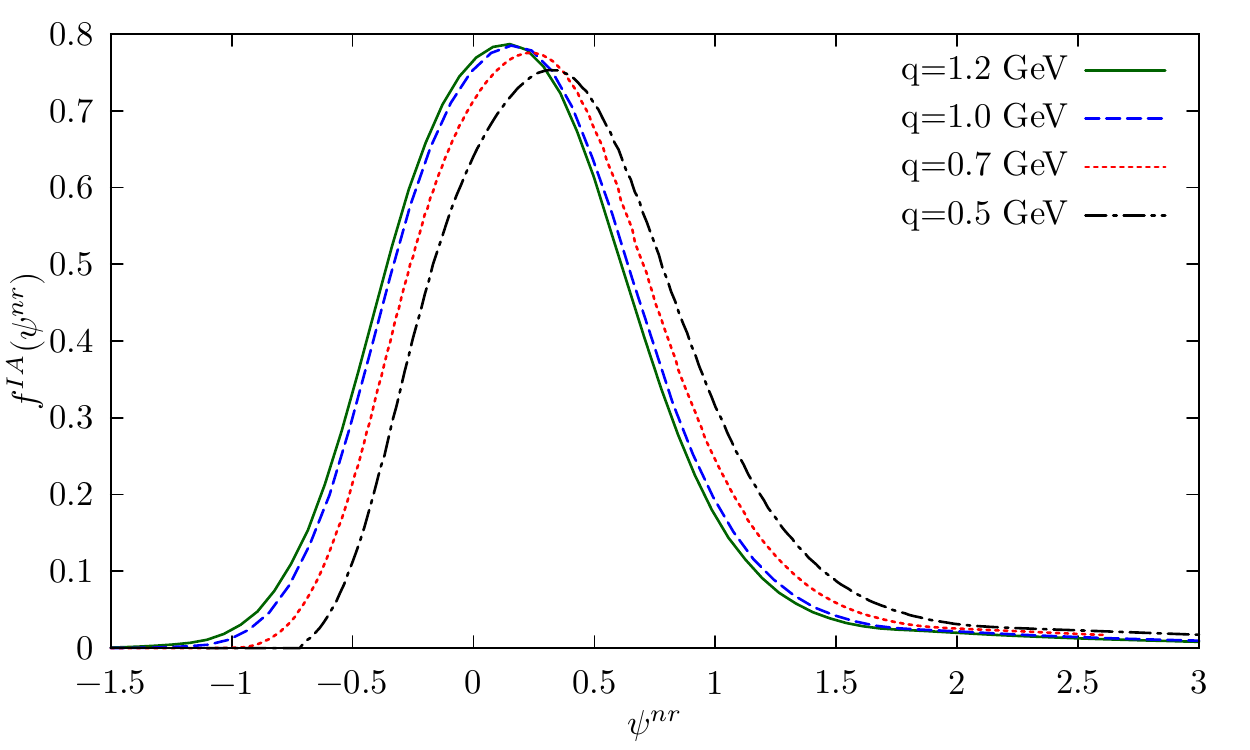} 
\caption{Scaling functions for $^{12}$C obtained in the IA from Eq.~\eqref{eq:fpwia} using the CBF hole SF for $|{\bf q}|=0.5,\, 0.7,\,1.0$, and 1.2 GeV. }
\label{f_sf_tot}
\end{figure}
The hole SF $\bar P_h({\bf p},E)$ is a function of two independent variables, which are related in a non trivial way. It is long known that the PWIA of Eq.~\eqref{eq:onebody-appr}, which disregards the dependence on the removal energy of the nucleus, is inaccurate. Realistic $P_h({\bf p},E)$ exhibits a strong correlation between momentum and removal energy, implying that large momenta always correspond to large removal energies. For instance, for nuclear matter hole SF calculated within the CBF approach,  around $50\%$ of the strength at $|\mathbf{p}|= 3$ fm$^{-1}$ resides at $E>200$ MeV~\cite{Benhar:1989aw}. Furthermore, the shell structure of the nucleus is completely disregarded in the PWIA of Eq.~\eqref{eq:onebody-appr}. 

In the following, we argue that the use of a realistic hole SF produces noticeably different scaling features of the nucleon-density response from those obtained within the PWIA model. In the IA, the energy conserving $\delta$ function of Eq.~\eqref{eq:dens_ia} reads
\begin{align}
\delta\left(\omega + E - \frac{|{\bf p}|^2}{2m} - \frac{|{\bf q}|^2}{2m} -\frac{|{\bf p}||{\bf q}|\cos\theta}{m}\right)\ .\label{eq:delta}
\end{align}
Imposing $|\cos\theta|\leq 1$ gives a boundary condition on both $E$ and $|{\bf p}|$, which are related through $\bar{P}_h({\bf p},E)$.  The Jacobian still yields a factor $\kappa$ that cancels the one of Eq.~\eqref{eq:f_pwia}. The binding energies associated to the continuum part of the hole SF are generally larger than $|{\bf p}|^2/2m + \mathcal{U}(|{\bf p}|)$.  This feature is particularly relevant for $\psi^{nr} > 1$, as larger values of $\omega$ are needed to compensate for the large removal energy. Hence, for sufficiently large momentum transfers we expect  violations of first-kind scaling, and the appearance of a more significant tail at the right of the quasielastic peak that will enhance the asymmetry of the scaling function compared to the PWIA case.  

Scaling violations are apparent in Fig.~\ref{f_sf_tot}, as the positions of the peaks of the scaling functions depend upon the momentum transfer. These shifts are likely to be ascribed to the energy of the bound hole state described by the hole SF, analogously to Fig.~\ref{scal_NM_in}. However,  the scaling functions obtained using the hole SF show a more pronounced asymmetric shape than those displayed in Fig.~\ref{scal_NM_in}.

\begin{figure}[]
\centering
\includegraphics[scale=0.675]{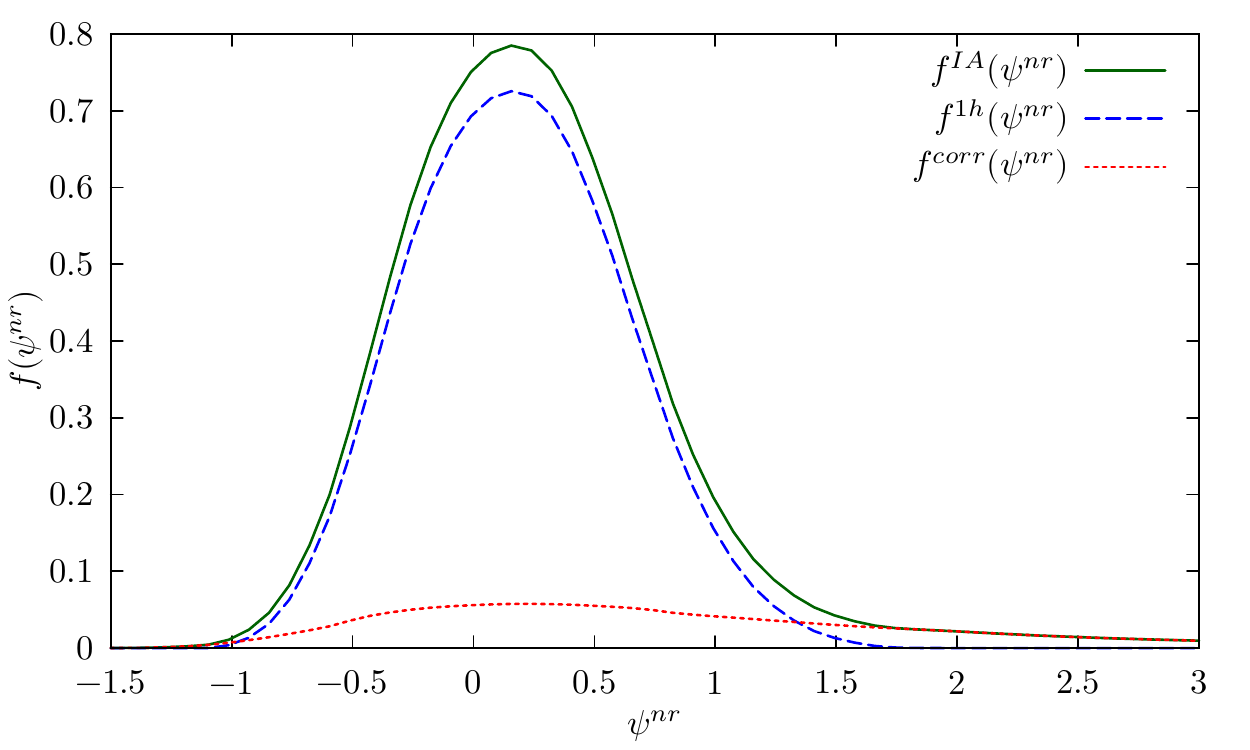}
\includegraphics[scale=0.675]{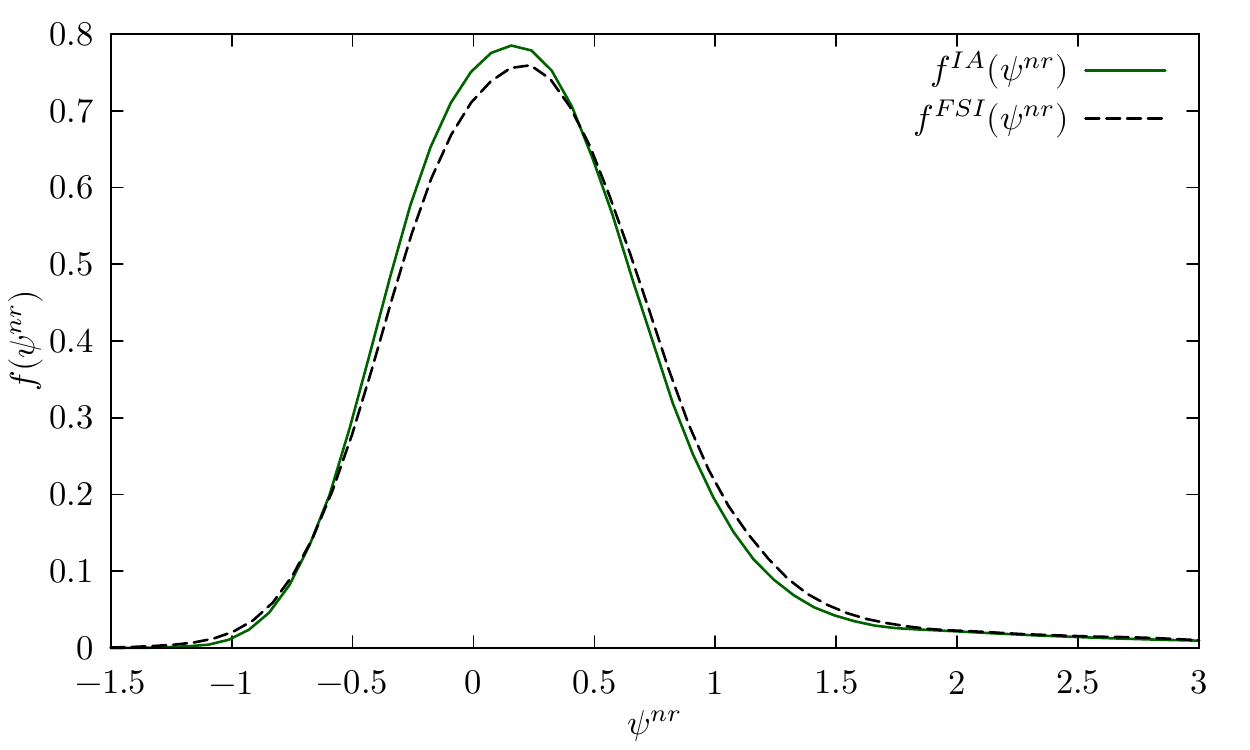}
\caption{Top: Breakdown of the scaling response of $^{12}$C at $|{\bf q}|$=1.2 GeV showed in Fig.~\ref{f_sf_tot} into the total, hole, and  background contributions. 
Bottom: Dashed (black) and solid (green) lines correspond to the scaling function calculated with and without the inclusion of FSI effect at $|{\bf q}|= 1$ GeV in $^{12}$C. The IA curve corresponds to that displayed in Fig.~\ref{f_sf_tot}, and it is used in the convolution detailed in Eq.~\eqref{dens:resp:folding} to incorporate the FSI effects. }
\label{f_fsi_NOfsi}
\end{figure}

In the upper panel of Fig.\ref{f_fsi_NOfsi} we show the breakdown of the scaling response at $|{\bf q}|$=1 GeV into the one-hole and  correlation contributions, coming from the pole and the continuum part of the hole Green's function. The asymmetric shape is mostly determined by the background contribution, with a large tail in the region of large $\psi$. Interestingly, the scaling response obtained by retaining only the one-hole contribution in the SF is not completely symmetric. This has to be ascribed to the presence of two independent integration variables, {\em i.e.} $|{\bf p}|$ and $E$. This more sophisticated description of nuclear dynamics likely contributes to the asymmetry observed in the experimental data. In the lower panel of the figure, we show how FSI affect the scaling function for $|{\bf q}|=\ 1$ GeV, comparing the IA (dashed black) and the total (solid green) results. Although FSI are significant for moderate momentum transfer, they are practically negligible in the kinematical region displayed in the figure. Overall FSI provide a shift and a redistribution of the strength of the scaling function, bringing about an enhancement of the asymmetry.

\section{Conclusions}
\label{sec:conclusions}

We have studied the scaling properties of the nucleon-density response, a key quantity to understand the scaling of the electromagnetic longitudinal and transverse response functions~\cite{Rocco:2017hmh}. The nucleon-density response of $^{12}$C has been calculated in the kinematical region in which collective excitations can be safely neglected. To this aim, we employed particle and hole SFs obtained within two many-body methods, both widely used to describe electroweak reactions in nuclei. 

We first consider the semi-phenomenological model developed in Ref.~\cite{FernandezdeCordoba:1991wf} and successfully applied to study a number of inclusive electro-weak reactions \cite{Nieves:2004wx, Nieves:2017lij,Marco:1995vb,Gil:1997bm,Oset:1994vp,Marco:1997xb,SajjadAthar:2007bz,SajjadAthar:2009cr}. This model relies on realistic particle and hole self-energies computed in isospin-symmetric nuclear matter and predictions for finite nuclei are made employing the LDA. Short-range effects are accounted for by an in-medium effective NN interaction. The latter, derived from the experimental elastic NN cross section, also incorporates some medium-polarization corrections through the RPA. The other approach, successfully tested in electroweak-nuclear reactions~\cite{Ankowski:2014yfa, Benhar:2005dj, Benhar:2006nr, Benhar:2009wi, Benhar:2010nx, Vagnoni:2017hll}, is based on a microscopic calculation of the hole SF, carried out within the CBF theory. The interaction of the relativistic struck nucleon with the spectator system is included via a convolution scheme, devised from a generalization of the Glauber theory describing high energy proton-nucleus scattering. 

We have shown that both approaches lead to compatible $^{12}$C nucleon-density scaling functions, characterized by an asymmetric shape, although less pronounced than the one of the experimental data. Whilst the CBF SF provides a more accurate description of the ground-state of $^{12}$C, presently it can only be applied to closed-shell nuclei. On the other hand, the LDA model can be readily extended to the $^{40}$Ar nucleus, which will be employed in future neutrino-oscillation experiments~\cite{dune_web}.

Employing a simplified model of nuclear dynamics, which retains the main aspects of the many-body problem, we discussed the dynamical origin of the scaling of the first kind exhibited by the nucleon-density response function.  We have argued that its asymmetric shape is mostly due to the $2h1p$ dynamics incorporated in the continuum component of the hole SF of Ref.~\cite{Benhar:1994hw}, that in turn accounts for NN correlations. On the other hand, the asymmetry is only slightly enhanced by FSI effects. The latter, relevant in the low momentum-transfer region only, lead to a shift of the peak position towards smaller values of $\psi^{nr}$ and to a redistribution of the strength towards larger values of $\psi^{nr}$. According to the relativistic mean field study carried out in Ref.~\cite{Caballero:2007tz}, the asymmetry of scaling function has to be ascribed to the dynamical enhancement of the lower component of the Dirac spinors, which are not present the non relativistic nucleon-density response function. Analogously to the GFMC results of Ref.~\cite{Rocco:2017hmh}, the asymmetry is also observed within the non relativistic scheme of nuclear dynamics based on the particle and hole SFs. Our results do not necessarily invalidate the relativistic mean field picture of scaling. The intriguing hypothesis that some of the non relativistic correlations might arise from a non relativistic reduction performed already at the mean filed level deserves further investigations. 

Within the SF formalism, we found that, once the pre-factors describing the single-nucleon interaction-vertices are divided out, the longitudinal and transverse electromagnetic response functions share a common kernel, intimately connected to the one of the nucleon-density response function. Consequently, the electromagnetic longitudinal and transverse scaling functions are very similar to the nucleon-density scaling function--the small differences being ascribable to discrepancies between GRFG and SF pre-factors. Therefore, besides two-body current and collective corrections effects, the breaking of zeroth and first kind scalings has be attributed to deficiencies in the nuclear model used to estimate the single-nucleon electroweak matrix elements in nuclei. 

\begin{acknowledgments}
 This research has been supported by the Spanish Ministerio de Econom\'\i a y
 Competitividad and European FEDER funds under  contracts FIS2014-51948-C2-1-P, FIS2014-51948-C2-2-P, FIS2017-84038-C2-1-P  and SEV-2014-0398, by
 Generalitat Valenciana under Contract PROMETEOII/2014/0068, by the U.S. Department of Energy, Office of Science, Office of Nuclear Physics, under 
 contract DE-AC02-06CH11357, and by the Centro Nazionale delle Ricerche (CNR) and the Royal Society under the CNR-Royal Society International Fellowship scheme 
 NF161046. Under an award of computer time provided by the INCITE program, this research used resources of the Argonne Leadership 
 Computing Facility at Argonne National Laboratory, which is supported by the Office of Science of the U.S. Department of Energy under contract DE-AC02-06CH11357.
 \end{acknowledgments}

\bibliography{biblio}

\end{document}